\pdfoutput=1

\documentclass[a4paper]{jpconf}
\usepackage{graphicx}
\newcommand{\cPT}{{\cal PT}}
\newcommand{\cP}{{\cal P}}
\newcommand{\cT}{{\cal T}}

\newcommand{\vep}{\varepsilon}
\newcommand{\half}{\textstyle{\frac{1}{2}}}

\begin{document}
\title{$\cPT$-symmetric classical mechanics}
\author{Carl M Bender${}^1$ and Daniel W Hook${}^{1,2}$}
\address{${}^1$Physics Department, Washington University, St. Louis, MO 63130,
USA}
\ead{cmb@wustl.edu}
\vspace{.2cm}
\address{${}^2$Centre for Complexity Science, Imperial College London, London
SW7 2AZ, UK}
\ead{d.hook@imperial.ac.uk}

\begin{abstract}
This paper reports the results of an ongoing in-depth analysis of the classical
trajectories of the class of non-Hermitian $\cPT$-symmetric Hamiltonians $H=p^2+
x^2(ix)^\vep$ ($\vep\geq0$). A variety of phenomena, heretofore overlooked, have
been discovered such as the existence of infinitely many separatrix
trajectories, sequences of critical initial values associated with limiting
classical orbits, regions of broken $\cPT$-symmetric classical trajectories, and
a remarkable topological transition at $\vep=2$. This investigation is a work in
progress and it is not complete; many features of complex trajectories are still
under study.
\end{abstract}

\section{Introduction}\label{s1}
The family of non-Hermitian $\cPT$-symmetric quantum-mechanical Hamiltonians
\begin{equation}
\hat{H}=\hat{p}^2+\hat{x}^2(i\hat{x})^\vep\quad(\vep~{\rm real})
\label{e1}
\end{equation}
is a complex deformation in the parameter $\vep$ of the quantum 
harmonic-oscillator Hamiltonian $\hat{p}^2+\hat{x}^2$. At $\vep=0$ the spectrum
of $\hat{H}$ consists of the harmonic-oscillator eigenvalues $E_n=2n+1$ ($n=0,\,
1,\,2,\,\dots$). As $\vep$ varies away from $0$, the eigenvalues exhibit several
characteristic behaviors: For $\vep>0$ the eigenvalues are all real, positive,
and discrete, and the eigenvalues increase as $\vep$ increases; for $\vep<0$
only a finite number of the eigenvalues are real and the infinite number of 
remaining complex eigenvalues occur as complex-conjugate pairs
\cite{R1,R2,R3,R4,R5,R6}.

The parametric region $\vep\geq0$ is called the {\it region of unbroken $\cPT$
symmetry} because all of the eigenfunctions of $\hat{H}$ are simultaneously
eigenstates of the $\cPT$ operator; the parametric region $\vep<0$ is called the
{\it region of broken $\cPT$ symmetry} because the eigenfunctions of $\hat{H}$
are not all eigenstates of $\cPT$ \cite{R6}. A transition at $\vep=0$ separates
the regions of unbroken and broken $\cPT$ symmetry. Typically, $\cPT$-symmetric
quantum Hamiltonians exhibit a transition with only real eigenvalues on one side
and partly real and partly complex eigenvalues on the other. For $\vep>0$ there
exists a formal similarity transformation from the non-Hermitian Hamiltonian
$\hat{H}$ to a Hermitian Hamiltonian $\hat{h}$, which is {\it isospectral} to
(has the same eigenvalues as) $\hat{H}$: $\hat{h}=S\hat{H}S^{-1}$. However, this
isospectral equivalence is only formal because $S$ and $S^{-1}$ are unbounded
operators, and thus the vectors in the domains of $\hat{H}$ and $\hat{h}$ are
not in one-to-one correspondence \cite{R6}.

The quartic case $\vep=2$ is special because only for this value of $\vep$ can
the transformation from $\hat{H}$ to $\hat{h}$ be carried out in closed form
with $\hat{h}$ having the simple local structure $\hat{p}^2+V(\hat{x})$, where
the potential function $V$ is also quartic \cite{R7,R8,R9}. For other values
of $\vep$ the corresponding isospectral Hermitian Hamiltonian $\hat{h}$ contains
$\hat{p}^2,\,\hat{p}^4,\,\hat{p}^6,\,\ldots$, and thus is nonlocal \cite{R10}.

\begin{figure}[h!]
\begin{center}
\includegraphics[scale=.22]{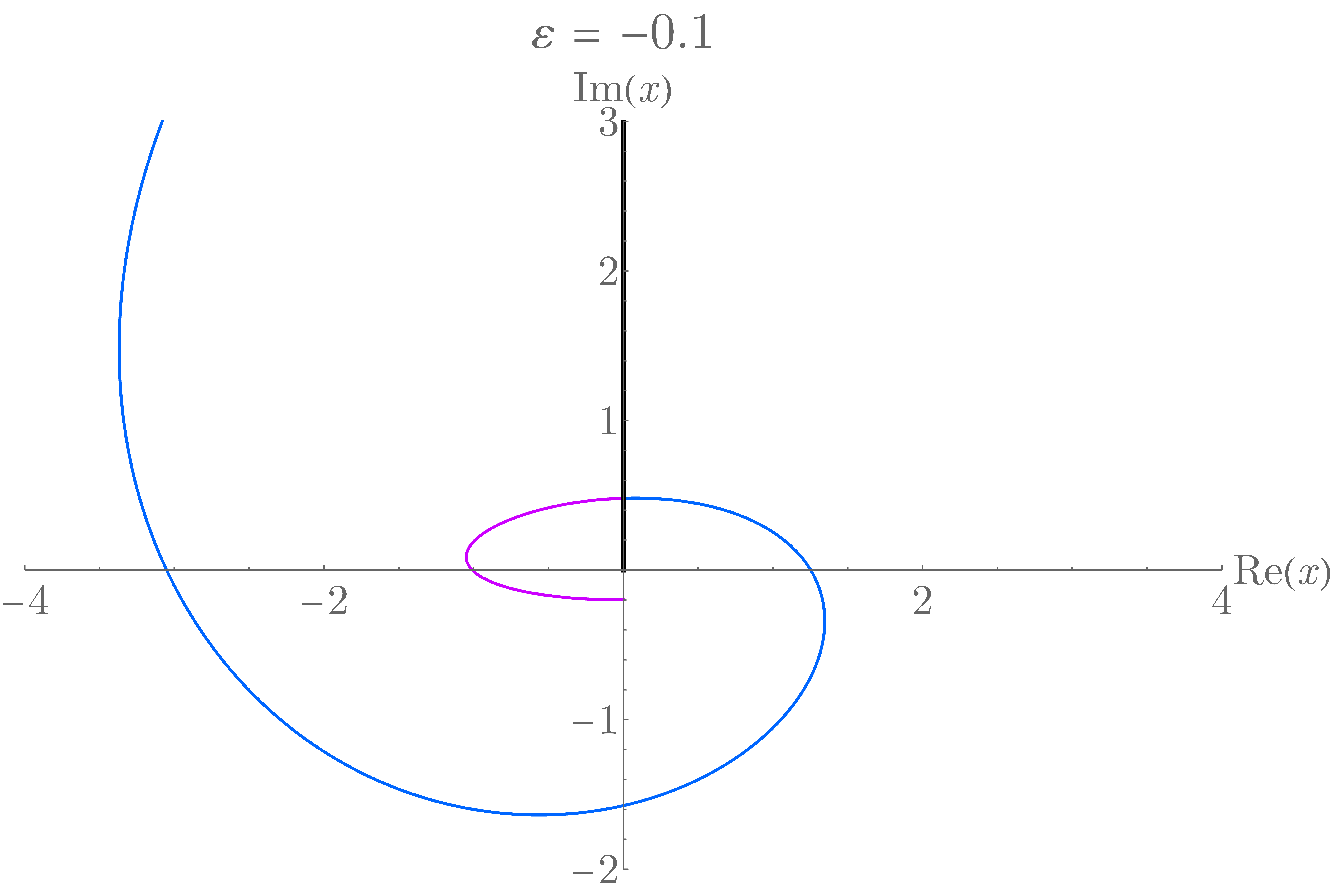}
\hspace{0.05cm}
\includegraphics[scale=.22]{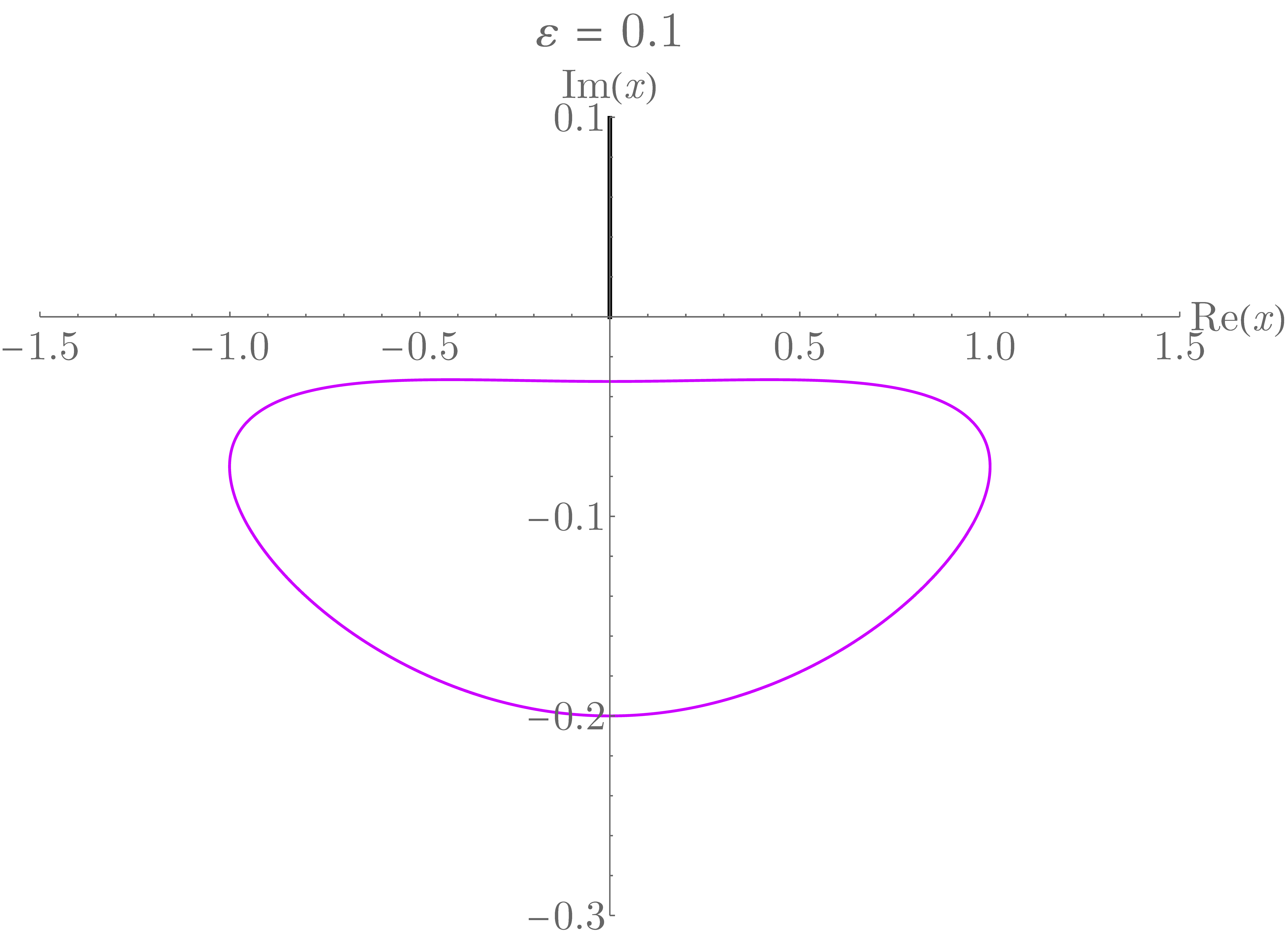}
\end{center}
\caption{[Color online] Complex classical trajectories for the Hamiltonian
$H=p^2+x^2(ix)^\vep$ for $\vep=-0.1$ and $\vep=0.1$. The energy (the value of
$H$) for the classical orbits is $1$ and the initial value in both cases is
$-0.2\,i$. For negative $\vep$ the orbits are open (left panel) and for positive
$\vep$ the orbits are usually closed (right panel).}
\label{f1}
\end{figure}

This paper reports our study of the remarkable underlying {\it classical}
properties of the quantum Hamiltonian (\ref{e1}). The classical trajectories are
curves $x(t)$ that lie in the complex-$x$ plane. At the classical level there is
a transition at $\vep=0$ that corresponds with the quantum transition described
above. Specifically, in the region of unbroken $\cPT$ symmetry the classical
orbits tend to be closed but in the broken region the orbits are open
\cite{R11}. This change of topology is shown in Fig.~\ref{f1} and the classical
topological transition between regions of broken and unbroken $\cPT$ symmetry
has been observed repeatedly in laboratory experiments
\cite{R12,R13,R14,R15,R16,R17,R18,R19,R20,R21,R22}.

There have already been many studies of complex classical mechanics
\cite{R2,Z1,Z2,Z3,Z4,Z5,Z6} and studies of complex quantum-mechanical systems at
the classical level \cite{Z7,Z8,Z9}. Here, we have undertaken a detailed
re-examination of the classical trajectories of $H$ in (\ref{e1}) in the
unbroken region $\vep\geq0$. We report here that the classical trajectories
exhibit many surprising properties that to date have not been noticed. For
example, there are sequences of critical initial values that give rise to
separatrix trajectories; these separatrices serve as boundaries between regions
of topologically distinct classes of orbits. A particularly interesting feature
is that at $\vep=2$, the unique value of $\vep$ at which there exists a simple
transformation from a quantum $\cPT$-symmetric non-Hermitian quantum Hamiltonian
to a local Hermitian quantum Hamiltonian, there is a corresponding dramatic
transition in the topology of the classical orbits. We give a detailed
description of the current status of this work in this paper.

This presentation is organized as follows. Section~\ref{s2} reviews the most
well known features of the complex classical trajectories for the Hamiltonian
(\ref{e1}). We examine the generic properties of complex classical trajectories
in the parametric region $0\leq\vep<2$ in Sec.~\ref{s3} with a focus on the case
of irrational $\vep$. (The case of rational $\vep$ is more complicated because
when $\vep$ is rational there are new discrete symmetries in addition to $\cPT$
symmetry that one must consider, and this complicates the discussion. We will
discuss the case of rational $\vep$ at length in a future paper.) We discuss the
properties of complex trajectories in region $0\vep\geq2$ in Sec.~\ref{s4}. Some
concluding remarks and conjectures about results of future investigations are
given in Sec.~\ref{s5}. 

\section{General properties of complex classical trajectories}\label{s2}
We begin by reviewing some of the properties of complex classical trajectories
for the Hamiltonian (\ref{e1}). Hamilton's classical equations of motion are
\begin{equation}
\dot{x}=\textstyle{\frac{\partial}{\partial p}}H=2p,\qquad
\quad\dot{p}=-\textstyle{\frac{\partial}{\partial x}}H=-(2+\vep)x(ix)^\vep,
\label{e2}
\end{equation}
where the dot indicates differentiation with respect to time $t$. (The time
variable $t$ is treated as real.) As is the case with any Hamiltonian system, an
immediate consequence of these equations is that the energy (that is, the value
of $H$) is conserved in time. For any $\vep$ we may perform a scaling of $x$ and
$p$ in the Hamiltonian (\ref{e1}) so that without loss of generality the
dimensionless total energy of a classical particle can be taken to be $1$:
\begin{equation}
p^2+x^2(ix)^\vep=1.
\label{e3}
\end{equation}

The system of first-order equations (\ref{e2}) is equivalent to Newton's
equation of motion, which is a second-order differential equation, but the fact
that energy is conserved allows us to use the first equation in (\ref{e2}) to
rewrite (\ref{e3}) as a {\it first-order} differential equation for the motion
of the classical particle:
\begin{equation}
\dot{x}^2+4x^2(ix)^\vep=4.
\label{e4}
\end{equation}
This equation emphasizes that if we know where the particle is, we know the
velocity of the particle up to a sign, and therefore we can predict where the
particle will go. (The choice in sign corresponds to whether the particle is
moving forward or backward in time.) Furthermore, because this equation is
complex, the path of the classical particle lies in the complex-$x$ plane.

In order to understand the trajectory of a classical particle it is necessary to
find the {\it classical turning points} of the Hamiltonian. At these points, the
velocity of the particle vanishes. From (\ref{e4}) we see that the turning
points of $H$ in (\ref{e1}) are solutions to
\begin{equation}
x^2(ix)^\vep=1,
\label{e5}
\end{equation}
so the turning points all lie on the unit circle in the complex-$x$ plane.

By solving (\ref{e4}) we find the classical particle trajectories $x(t)$
and we can easily establish that the period $T$ of any closed orbit $C$ is given
by \cite{R3}
\begin{equation}
T=\frac{1}{2}\oint_C\frac{dx}{\sqrt{1-x^2(ix)^\vep}}.
\label{e6}
\end{equation}
Typically, a closed path $C$ in the complex-$x$ plane (or, more generally, the
Riemann surface) encloses the square-root branch-cut singularity joining a
$\cPT$-symmetric pair of turning points, as we can see from (\ref{e6}). Cauchy's
theorem, which guarantees path independence of complex contour integrals,
implies that for any value of $\vep$ the period of all closed orbits that
enclose a given pair of turning points is the same.

\begin{figure}[h!]
\begin{center}
\includegraphics[scale=.47]{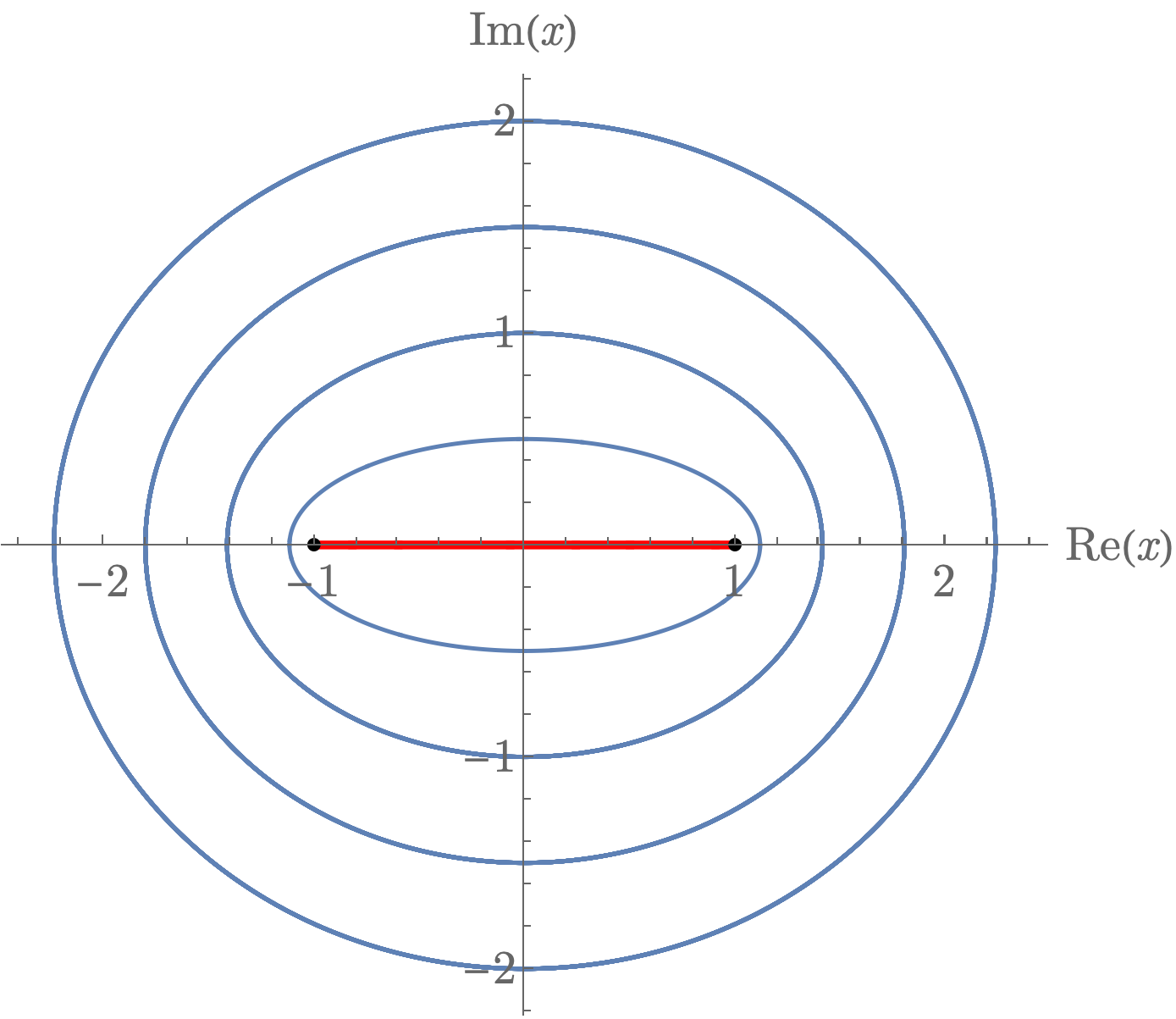}
\end{center}
\caption{[Color online] Five closed complex trajectories of the classical
harmonic oscillator $H=p^2+x^2$, which is the $\vep=0$ case of $H$ in
(\ref{e1}). The energy (the value of $H$) for all classical orbits is $1$. Four
orbits are nested ellipses whose foci are the turning points (black dots)
located at $x=\pm1$. These elliptical orbits pass through the points $-0.5\,i$,
$-i$, $-1.5\,i$, and $-2.0\,i$. A special singular orbit (bold red) lies on the
real axis and terminates at the turning points. The period of every orbit has
the same value $T=\pi$. Although the orbits are elliptical, they are unlike
Galilean planetary orbits, which are also nested ellipses, because the periods
of Galilean orbits increase as the size of the ellipses increases. (Jupiter's
year is longer than the Earth's year.) The Hamiltonian $H=p^2+x^2$ is both $\cP$
symmetric and $\cT$ symmetric, so all trajectories are left-right and also
up-down symmetric.}
\label{f2}
\end{figure}

For the elementary case $\vep=0$, where the classical trajectories are governed
by the classical harmonic oscillator Hamiltonian $H=p^2+x^2$, there are only two
turning points, which are located at $x=\pm1$. All classical trajectories
for this Hamiltonian are closed and periodic and lie in a (one-sheeted) complex
plane. Five such trajectories are shown in Fig.~\ref{f2}.

All but one of the trajectories in Fig.~\ref{f2} belong to the continuous family
of nested ellipses whose foci are the turning points; these ellipses fill the
entire complex-$x$ plane except for the real axis between the turning points. In
addition, there is one classical trajectory for a particle that oscillates
between the two turning points; this particle reaches a turning point in finite
time, stops, and then retraces its path as it moves away from the turning point.
We call this special trajectory, which representing the motion of a particle
that oscillates between a pair of turning points, a {\it terminating}
trajectory.

The terminating trajectory is {\it singular} because both ends of the trajectory
terminate at turning points while all other trajectories represent a particle
that {\it never} stops or reverses its direction. Note that in the limit as the
semiminor axis of an elliptical path approaches 0 we obtain a {\it degenerate}
ellipse that represents a particle that travels parallel to and infinitesimally
above (or below) the real axis, goes around a turning point infinitesimally
close to the turning point, and continues on again parallel to and
infinitesimally below (or above) the real axis. This limiting trajectory does
not terminate. The singular terminating trajectory lies {\it exactly} on the
real axis between the two turning points; the terminating trajectory is special
because it does not encircle and enclose the turning points.

To summarize: All elliptical classical orbits enclose the branch cut on the real
axis that joins the two turning points while the singular orbit that oscillates
between the turning points terminates at the turning points and does not enclose
the branch cut. All classical orbits, the elliptical orbits and the terminating
orbit, have period $T=\pi$, which we calculate by evaluating the integral
(\ref{e6}).

In general, the shape of a classical trajectory in the complex plane is
determined by the turning points in the vicinity of the trajectory. Recall what
happens in the neighborhood of a turning point on the real axis: The turning
points in Fig.~\ref{f2} mark the edges of the {\it classically allowed} region,
which is the portion of the real axis between the turning points. A particle in
the classically allowed region slows down as it approaches a turning point. The
particle stops momentarily at the turning point and then speeds up as it goes
back along the path and moves away from the turning point. This trajectory
follows (and is parallel to) the real axis. However, in the {\it classically
forbidden} regions on the real axis to the right of the right turning point and
to the left of the left turning point, all classical trajectories are {\it
orthogonal} to the real axis.

\begin{figure}
\begin{center}
\includegraphics[scale=0.33]{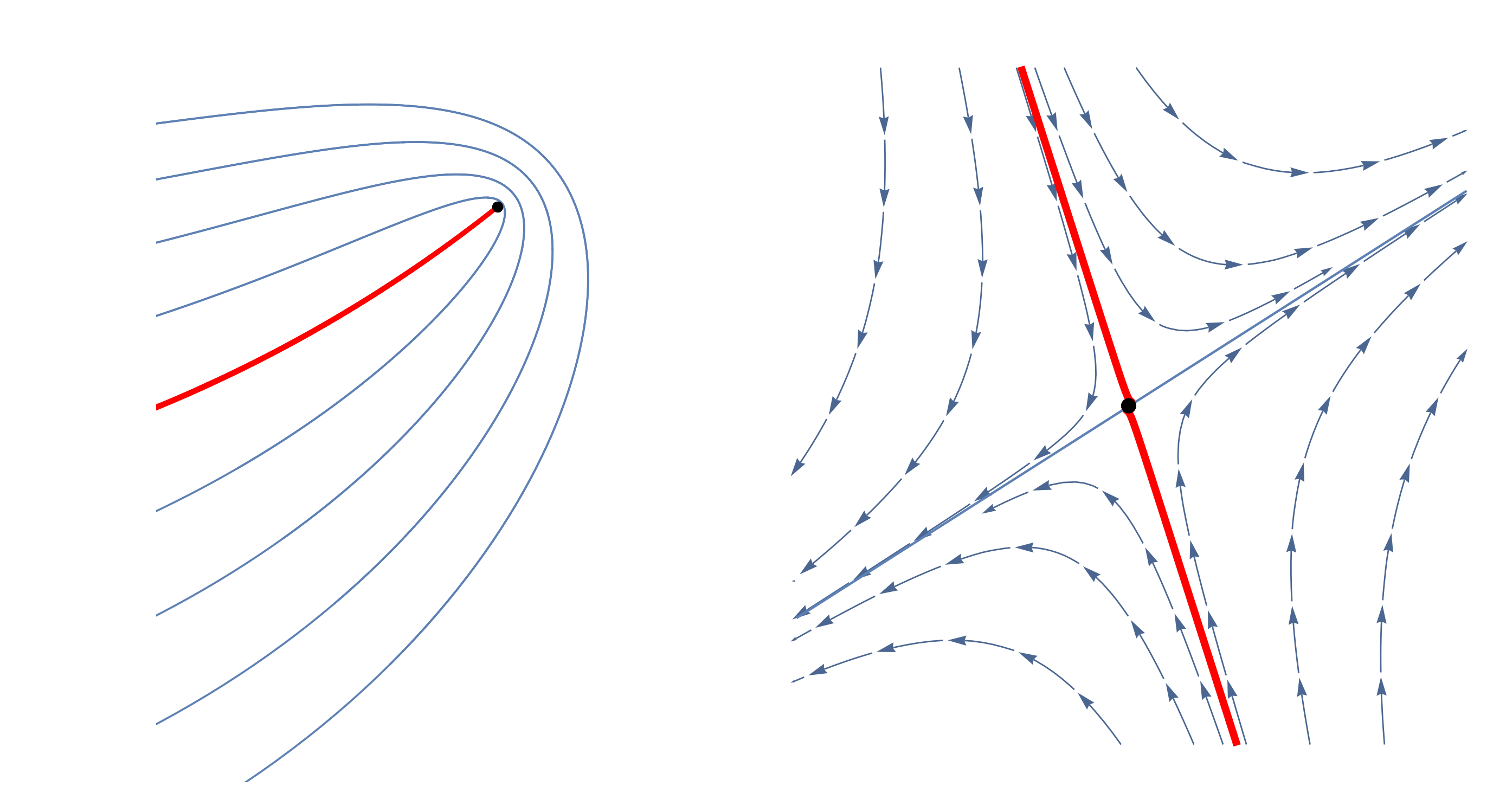}
\end{center}
\caption{[Color online] Schematic comparison of the behavior of trajectories of
a complex one-degree-of-freedom Hamiltonian near a typical turning point (black
dot) in the complex plane (left panel) and trajectories of a real
two-degree-of-freedom dynamical system near a typical saddle point (black dot)
(right panel). For the complex Hamiltonian there is a unique terminating path
(bold red) along which the trajectory approaches the turning point, stops, and
then moves away from the turning point. All other trajectories (blue) exhibit a
sling-shot-like behavior in which they make a $360^\circ$ turn as they go around
the turning point. For a saddle point there are two paths (bold red) along which
particles approach this critical point, but these particles cannot reach the
critical point in finite time. Along all other curves (blue) particles at first
approach the saddle point and then veer away from it.}
\label{f3}
\end{figure}

Let us extend this picture into the complex plane: In the neighborhood of a
turning point in the complex plane there is just one direction towards and away
from the turning point and there is a unique singular terminating curve along
which the classical particle (i) slows down as it approaches the turning point;
(ii) stops instantaneously at the turning point; and then (iii) speeds up as it
retraces this path and moves away from the turning point \cite{R9}. Particles
following other complex paths are flung around the turning point. The turning
point acts as if it is pulling on the classical particle; paths near the
terminating (in-and-out) trajectory make a sling-shot-like $360^\circ$ turn
about the turning point. Some typical classical trajectories are illustrated
schematically in Fig.~\ref{f3} (left panel). Note that the behavior illustrated
in this figure is specific to theories for which the kinetic-energy term in the
Hamiltonian is quadratic in $p$. For Hamiltonians whose kinetic-energy term is
$p^3$ or $p^4$, for example, the behavior of classical trajectories near the
turning points is significantly more complicated \cite{R9,R23}. We do not
discuss this possibility here.

Note that we are discussing here a Hamiltonian dynamical system having {\it one
complex degree of freedom}. Its classical trajectories are paths $x(t)$ in the
complex-$x$ plane. This is not the same as a system having two real degrees of
freedom. For the real system, the structure analogous to a complex turning point
is a critical point called a {\it simple saddle point}. Near a typical
(quadratic) saddle point [see Fig.~\ref{f3} (right panel)] there are {\it two}
terminating paths (red) directly towards the critical point. On these two
special curves the particle slows down as it approaches the critical point but
does not reach the critical point in finite time. On all other curves the
particle first slows down as it approaches the saddle point and then speeds up
as it turns and moves away from the saddle point. 

\begin{figure}
\begin{center}
\includegraphics[scale=0.29]{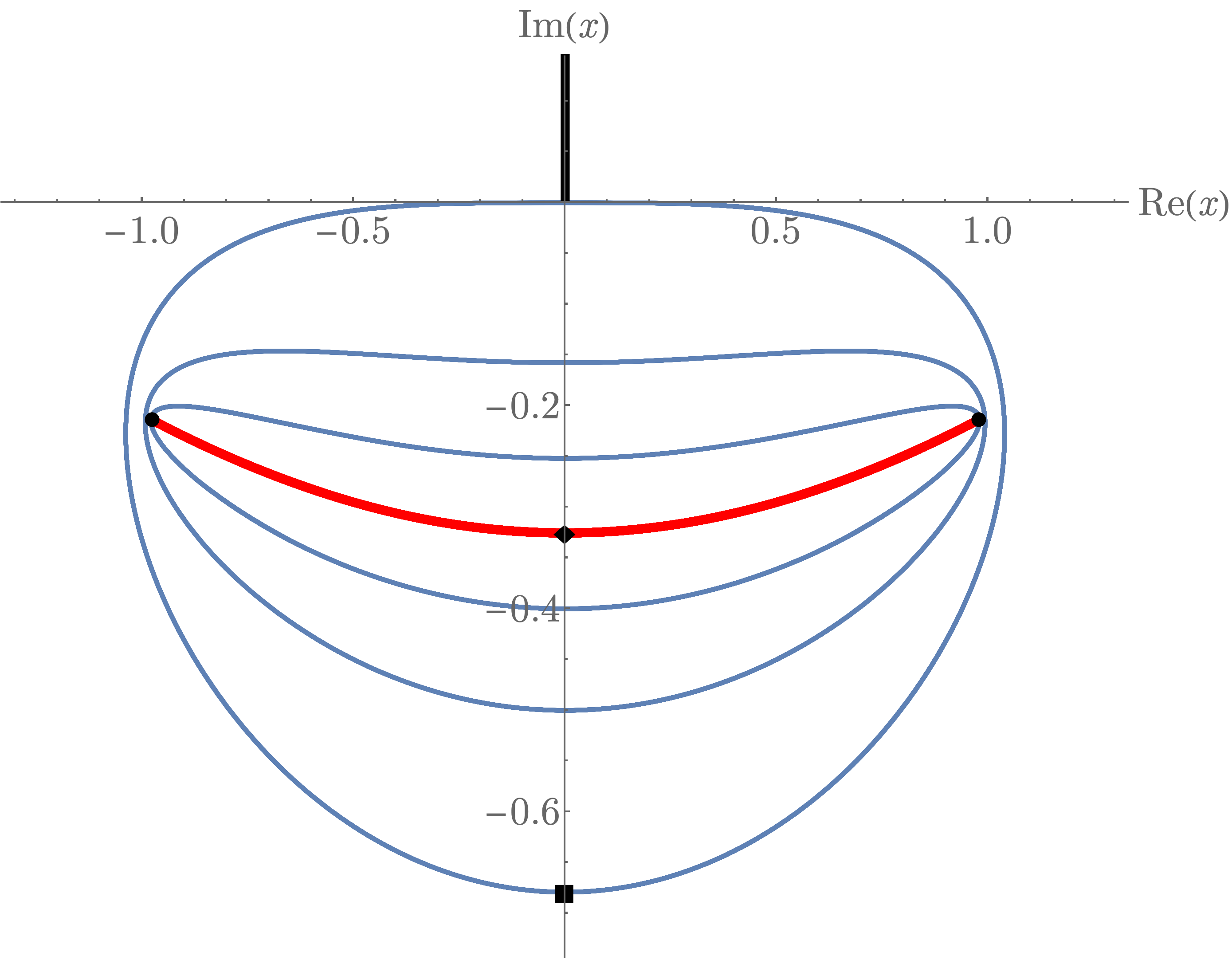}
\end{center}
\caption{Four complex classical trajectories of $H$ in (\ref{e1}) for $\vep=1/
\pi$. The energy is 1. A terminating singular trajectory (bold red) begins at
$s_0=-0.325235\,i$ (black diamond) and joins the two turning points (bold dots),
which are located at $t_0=\pm 0.88496-0.53483\,i$. Three more trajectories,
which begin at $-0.4\,i$, $-0.5\,i$, and $=-0.6438554\,i$ enclose the turning
points.  The point $x_0=-0.679076\,i$ (black square) is a {\it critical point}
because the trajectory that emerges from this point is the {\it last trajectory
that does not cross the branch cut} (black line on the positive-imaginary axis);
this trajectory is a {\it separatrix}. The periods for all closed periodic
orbits shown in this figure is $T=2.93702$. None of the trajectories shown cross
the branch cut but, as we will see, trajectories that begin below $x_0$ on the
imaginary axis cross the branch cut on the positive-imaginary axis at
successively higher points.}
\label{f4}
\end{figure}

As we increase $\vep$ in $H$ in (\ref{e1}) above $0$, the right and left turning
points shown on Fig.~\ref{f2} rotate downward into the lower-half complex plane
and are located at
\begin{equation}
x_{\rm right}=\exp\left(-\textstyle{\frac{i\pi\vep}{2\vep+4}}\right)\quad{\rm
and}\quad x_{\rm left}=\exp\left(-i\pi+\textstyle{\frac{i\pi\vep}{2\vep+4}}
\right),
\label{e7}
\end{equation}
which are $\cPT$ reflections of one another. [Under a $\cPT$ reflection a point
$x$ in the complex-$x$ plane goes into $-x^*$ \cite{R6}, and because $H$ in
(\ref{e1}) is $\cPT$ symmetric, its turning points are symmetric with respect to
reflection about the imaginary axis.] Because the potential becomes multivalued
when $\vep>0$, the equations of motion (\ref{e2}) and (\ref{e4}) of the
classical particle are defined on a Riemann surface. Thus, the complex-$x$ plane
in Fig.~\ref{f2} extends onto a multisheeted Riemann surface with a branch cut
emanating from the origin $x=0$. To be consistent with $\cPT$ symmetry we take
the branch cut to lie on the positive-imaginary axis (see Fig.~\ref{f4}).

We evaluate the integral (\ref{e6}) and find that the period of the closed
orbits is \cite{R3}
\begin{equation}
T=2\sqrt{\pi}\,\cos\left(\textstyle{\frac{\pi\vep}{4+2\vep}}\right)\Gamma\left(
\textstyle{\frac{3+\vep}{2+\vep}}\right)\big/\Gamma\left(\textstyle{\frac{4+
\vep}{4+2\vep}}\right).
\label{e8}
\end{equation}
The derivation of (\ref{e8}) assumes that the contour $C$ in (\ref{e6}) {\it
does not cross the branch cut on the positive-imaginary axis}.

Figure~\ref{f4} displays some classical trajectories for $\vep=1/\pi$. Plotted
are the terminating trajectory for a particle that oscillates between the
turning points, which lie at $t_0=\pm 0.84496-0.53483\,i$, and three more
trajectories that enclose this pair of turning points. These trajectories are
$\vep$-deformed versions of the $\vep=0$ ellipses in Fig.~\ref{f2}. No
trajectories shown in Fig.~\ref{f4} cross the branch cut. The periods for all of
the closed trajectories that do not cross the branch cut, as obtained from
(\ref{e8}) (and verified numerically), are $T=2.93702$.

\begin{figure}
\begin{center}
\includegraphics[scale=0.32]{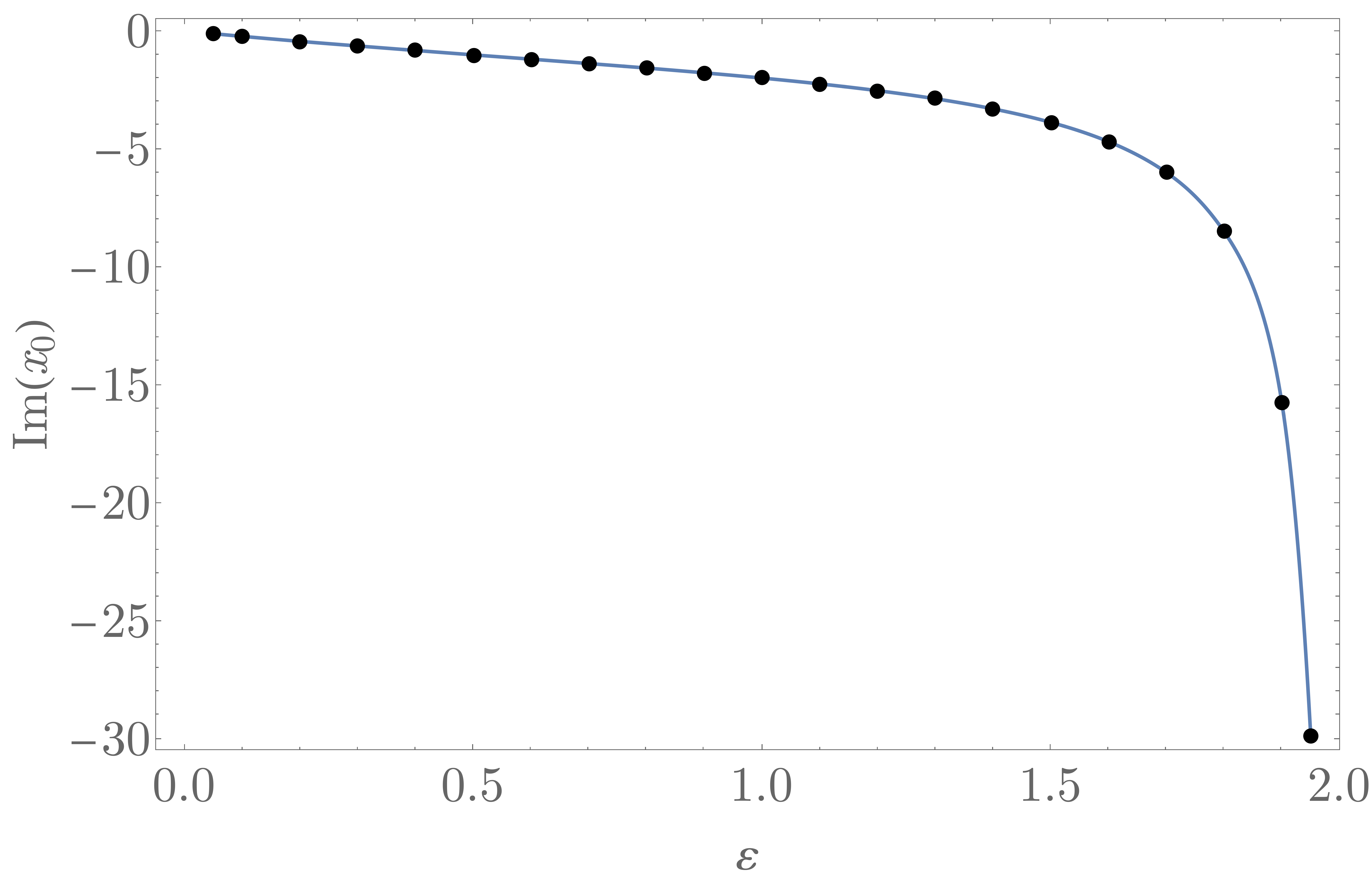}
\end{center}
\caption{Plot of the critical point $x_0$ (the lower point at which a closed
classical trajectory crosses the negative-imaginary axis for which the
corresponding upper crossing point is the origin) as a function of $\vep$. As
$\vep$ approaches $1.0$, $x_0$ approaches $-2.0\,i$ and as $\vep$ approaches
$2.0$, $x_0$ approaches $-i\infty$. The trajectory that originates from the
critical point is a {\it separatrix} that serves as the boundary curve between
orbits that remain on the principal sheet and orbits that cross the branch cut
and visit other sheets of the Riemann surface.}
\label{f5}
\end{figure}

The maximal trajectory (the largest trajectory in Fig.~\ref{f4} that does {\it
not} cross the branch cut on the positive-imaginary axis) begins at the critical
point $x_0=-0.6438554\,i$. This trajectory is a separatrix curve that serves as
the boundary between (i) trajectories that begin above $x_0$ and remain on the
principal sheet of the Riemann surface, and (ii) trajectories that begin below
$x_0$, cross the branch cut, and enter other sheets of the Riemann surface. As
$\vep$ increases, $x_0$ moves down the imaginary axis. As $\vep$ approaches
$1.0$, $x_0$ approaches $-2.0\,i$ and as $\vep$ approaches $2.0$, $x_0$
approaches $-i\infty$ (see Fig.~\ref{f5}). Thus, {\it there is a critical
topological transition in the classical theory at} $\vep=2$. This is a new
discovery.

\begin{figure}
\begin{center}
\includegraphics[scale=0.32]{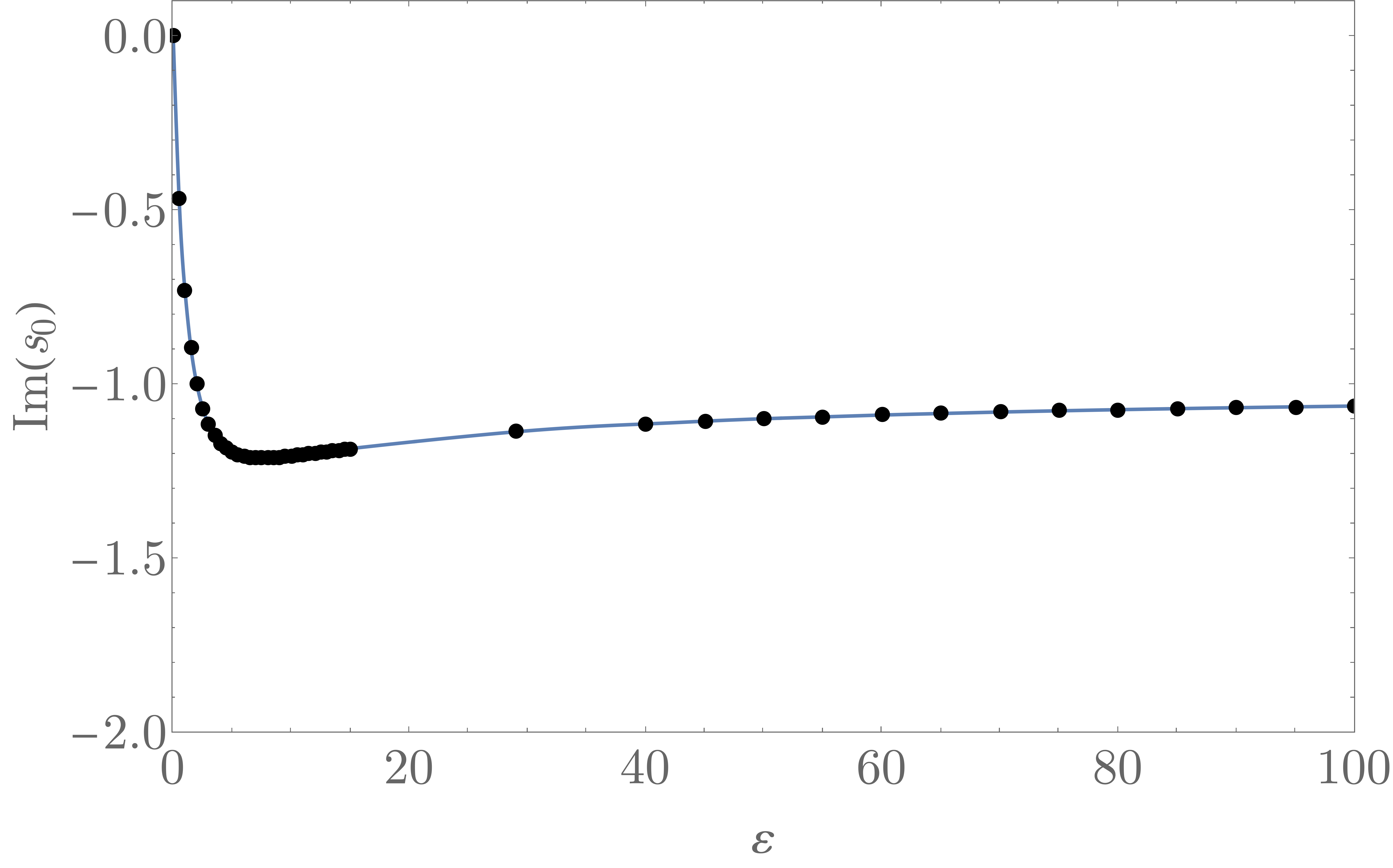}
\end{center}
\caption{Plot of the special initial value $s_0$ that gives rise to the
terminating trajectory as a function of $\vep$. Note that $s_0=-i$ at $\vep=2$.
As $\vep$ continues to increase, $s_0$ passes through a minimum value of
$-1.21188\,i$ at $\vep=7.62547$ and then approaches the limiting value $-i$ as
$\vep\to\infty$.}
\label{f6}
\end{figure}

Between the origin and the critical point $x_0$, where the separatrix trajectory
crosses the negative-imaginary axis, is the special point $s_0$. The point $s_0$
lies on the negative imaginary axis and is the center of the terminating
trajectory connecting the turning points $t_0$. Like the plot of $x_0$ in
Fig.~\ref{f5}, Fig.~\ref{f6} shows the value of $s_0$ as a function of $\vep$.
Unlike $x_0$, $s_0$ remains {\it finite} for all $\vep$. Observe that $s_0$
attains the value $-i$ when $\vep=2$. Then, as $\vep$ increases further, $s_0$
reaches a minimum value of about $-1.21188\,i$ at $\vep=7.62547$ and then rises
to $-i$ as $\vep\to\infty$.

It is easy to explain this asymptotic limit. We see from (\ref{e7}) that the
turning points approach $-i$ on the negative-imaginary axis as $\vep\to\infty$.
Furthermore, we see from (\ref{e2}) that a trajectory that crosses the imaginary
axis at the point $-iL$, where $L>0$, has a {\it real} velocity of $\pm\sqrt
{1+L^{\vep+2}}$, and thus the trajectory is {\it perpendicular} to this axis.
(The reality of the velocity is a consequence of $\cPT$ symmetry.) We conclude
that the terminating trajectory becomes infinitely short and crosses the
imaginary axis at $-i$ as $\vep$ approaches $\infty$. The triviality of this
trajectory at $\vep=\infty$ corresponds to a dramatic simplification in the
corresponding quantum-mechanical theory, which becomes exactly solvable (the
complex equivalent of a square-well potential) for infinite $\vep$ \cite{R24}.

A feature of the terminating curve that one can calculate analytically is the
{\it slope} of the terminating curve at a turning point. To determine this slope
we use Newton's law to find the acceleration $\ddot x$ of a classical particle
at the turning point. To derive Newton's law we take the time derivative of the
first equation in (\ref{e2}) and use the second equation in (\ref{e2}) to
eliminate $\dot p$:
\begin{equation}
\ddot x=-2(2+\vep)x(ix)^\vep.
\label{e9}
\end{equation}
This equation gives the (complex) acceleration $\ddot x$ as a function of the
position of the classical particle. If we substitute the location of the right
turning point $x_{\rm right}$ in (\ref{e7}) for $x$ in (\ref{e9}), we obtain \begin{equation}
\ddot x=-(4+2\vep)\exp\big(i\textstyle{\frac{\pi\vep}{4+2\vep}}\big).
\label{e10}
\end{equation}
Thus, at the right turning point the particle accelerates downward and to the
left and the terminating curve slopes upward at an angle of $\frac{\pi\vep}{4+2
\vep}$. For example, in Fig.~\ref{f4} the slope of the terminating curve is
$\frac{\pi}{2+4\pi}$, which we have verified numerically.

\section{Trajectories that leave the principal sheet}\label{s3}
The class of nested trajectories shown in Fig.~\ref{f4} densely fills the closed
region in the complex plane that is bounded by the critical trajectory passing
through the critical point $x_0$. We refer to his region as $R_0$, where the
subscript 0 indicates that $R_0$ contains the 0th pair of turning points $t_0$
that the terminating trajectory $s_0$ connects.

\begin{figure}
\begin{center}
\includegraphics[scale=0.27]{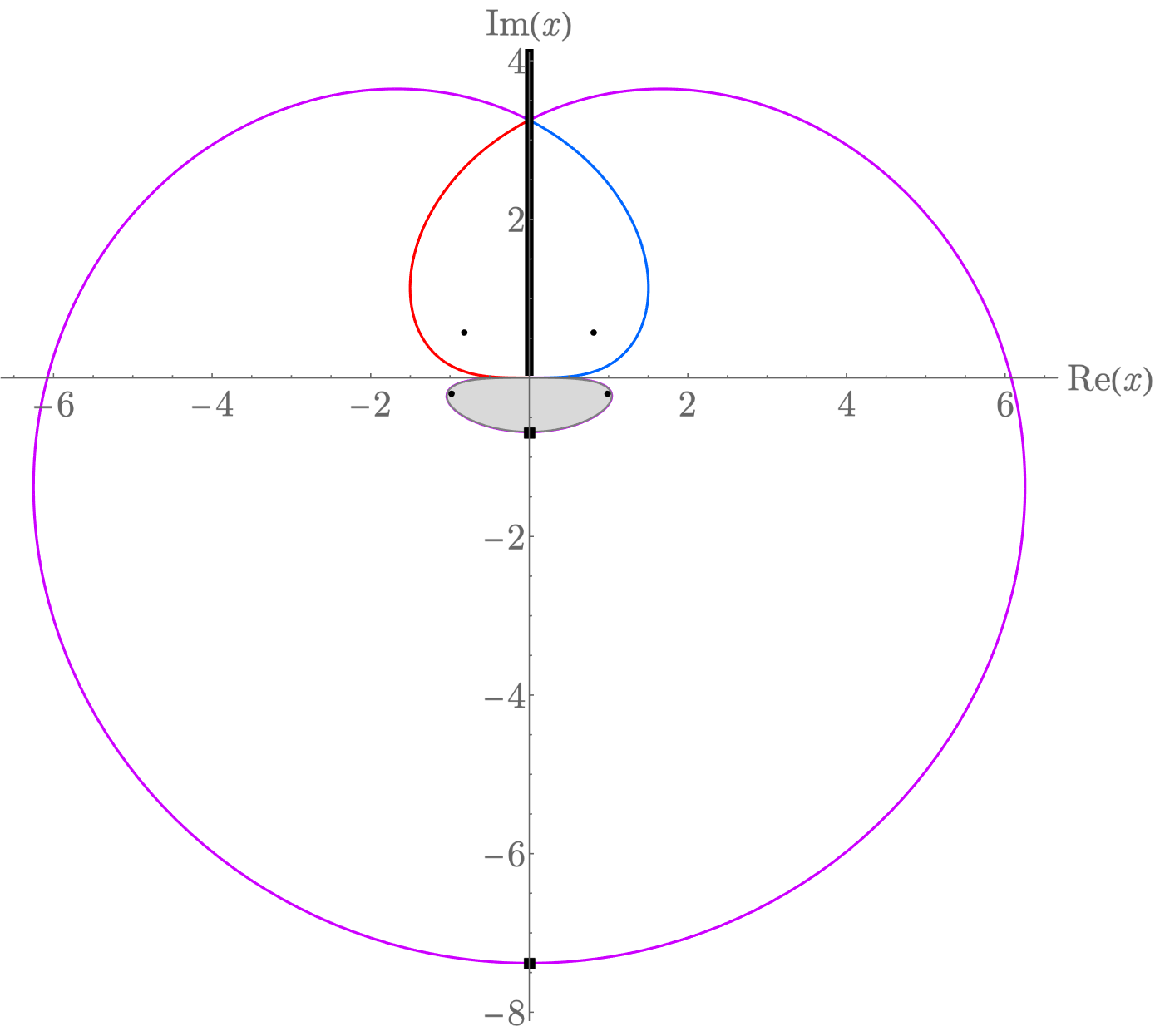}
\end{center}
\caption{[Color online] A trajectory that begins at $-0.68\,i$, which is just
below the critical point at $x_0=-0.679076\,i$ (black square) for $\vep=1/\pi$.
The trajectory can go right or left, but if it goes to the right, it follows the
edge of Region $R_0$ (gray) containing trajectories that lie entirely on the
principal sheet (sheet 0) of the Riemann surface. It then crosses the branch cut
(black line) on the positive-imaginary axis in the positive sense and enters the
sheet above the principal sheet (sheet 1, where the trajectory is colored red).
The trajectory then curves to the right and crosses back onto the principal
sheet, where it is colored purple. After making a full $360^\circ$ clockwise
turn in the principal sheet, it crosses the branch cut in the negative direction
and enters the sheet below the principal sheet (sheet $-1$, where the trajectory
is colored blue). It then turns around, crosses the branch cut in the positive
direction, returns to the principal sheet, and follows the edge of the gray
Region $R_0$ back to the starting point $-0.68\,i$ on the negative-imaginary
axis. This trajectory is $\cPT$ symmetric; that is, it is invariant under
reflection through the imaginary axis on sheet 0. The trajectory also crosses
the negative-imaginary axis just above the critical point $x_1=-7.389098\,i$
(black square), which marks the boundary between curves in Region $R_1$ and
curves in Region $R_2$. This closed trajectory encloses the {\it next} pair of
turning points (black dots), which are located at $t_1=\pm 0.81697+0.57668,i$ in
the upper-half complex planes on sheets 1 and $-1$, but does {\it not} enclose
the turning points at $t_0=\pm 0.97908-0.20346\,i$, which are also shown on
Fig.~\ref{f4}.}
\label{f7}
\end{figure}

A trajectory that begins slightly below the critical point $x_0=-0.679076\,i$ in
Fig.~\ref{f4} is shown in Fig.~\ref{f7}. This trajectory wraps around the edge
of the region of periodic trajectories in Fig.~\ref{f4} that are confined to the
principal sheet (sheet 0, where ${\rm arg}\,z$ ranges from $-\frac{3}{2}\pi$ to
$\half\pi$), and then briefly dips into the sheet above sheet 0 (sheet $1$,
where ${\rm arg}\,z$ ranges from $\half\pi$ to $\frac{5}{2}\pi$) and also the
sheet below sheet 0 (sheet $-1$, where ${\rm arg}\,z$ ranges from $-\frac{7}{2}
\pi$ to $-\frac{3}{2}\pi$).

The trajectory shown in Fig.~\ref{f7} is a separatrix that bounds a new region 
that we call $R_1$. The subscript 1 indicates that this region contains the pair
of turning points $t_1$, which are connected by the terminating trajectory
$s_1$. The full set of turning points are located at
$$\exp\Big[\textstyle{\frac{(4N-\vep)\pi i}{2\vep+4}}\Big]\quad(N=0,\,\pm1,\,
\pm2,\,...),$$
where the 0th pair of turning points $t_0$ is given by $N=0,\,-1$, the 1st pair
of turning points $t_1$ is given by $N=1,\,-2$, the 2nd pair of turning points
$t_2$ is given by $N=2,\,-3$, and so on.

To compute and plot trajectories like that in Fig.~\ref{f7}, which visit
multiple sheets of the Riemann surface, we rewrite Hamilton's equations in
(\ref{e2}) in terms of polar variables. We introduce polar representations for
$x(t)$ and $p(t)$:
$$x(t)=r(t)e^{i\theta(t)},\qquad p(t)=a(t)e^{i\alpha(t)},$$
where $r,\,\theta,\,a,\,\alpha$ are real and $r,\,\alpha\geq0$. We then take the
real and imaginary parts of the equations of motion (\ref{e2}) to obtain four
coupled real first-order differential equations:
\begin{eqnarray}
\dot{r}(t)&=&2a(t)\cos[\alpha(t)-\theta(t)],\nonumber\\
\dot{\theta}(t)&=&\textstyle{\frac{2a(t)}{r(t)}}\sin[\alpha(t)-\theta(t)],
\nonumber\\
\dot{a}(t)&=&(2+\vep)r^{\vep+1}(t)\cos[(1+\vep)\theta(t)-\alpha(t)+(\vep-2)
\pi/2],
\nonumber\\
\dot{\alpha}(t)&=&\textstyle{\frac{\vep+2}{a(t)}}r^{\vep+1}(t)\sin[(\vep+1)
\theta(t)-\alpha(t)+(\vep-2)\pi/2].
\label{e11}
\end{eqnarray}
The coupled system (\ref{e11}) is easy to solve numerically because it is first
order and we can use Runge-Kutta methods to do so. An advantage of using this
approach is that we can calculate the energy at each time step to verify that it
is conserved. This is a crucial diagnostic that we use to verify the accuracy of
our results.

The shape of a trajectory is controlled by turning points in the complex plane.
In general, turning points tend to {\it pull} on trajectories \cite{R23}. We see
in Fig.~\ref{f4} that there are two turning points $t_0$ on the principal sheet,
and from (\ref{e7}) we know that they are located in the lower-half plane at the
angles $-\frac{\pi}{ 2+4\pi}$ and at $-\pi+\frac{\pi}{2+4\pi}$. The angular
distance between these turning points is $\pi-\frac{\pi}{1+2\pi}$, and if we
rotate clockwise or anticlockwise by this angular amount from the turning points
$t_0$ on the principal sheet we find the next pair of turning points $t_1$, one
at $\frac{2\pi-1}{2+4\pi} \pi$ on sheet 1 and one at $-\frac{1+8\pi}{2+4\pi}\pi$
on sheet $-1$.

In general, even though there are many turning points that influence the shape
of a classical trajectory, a closed trajectory encloses (encircles) exactly two
turning points; that is, if we think of the trajectory as a directed closed
curve, then exactly two turning points lie in the {\it interior} of the region
delimited by the trajectory. This is a key property of a classical Hamiltonian
whose kinetic term is $p^2$, regardless of the number of Riemann sheets visited
by the trajectory. (Closed trajectories for a $p^3$ or $p^4$ Hamiltonian may
enclose more than two turning points.)

\begin{figure}
\begin{center}
\includegraphics[scale=0.54]{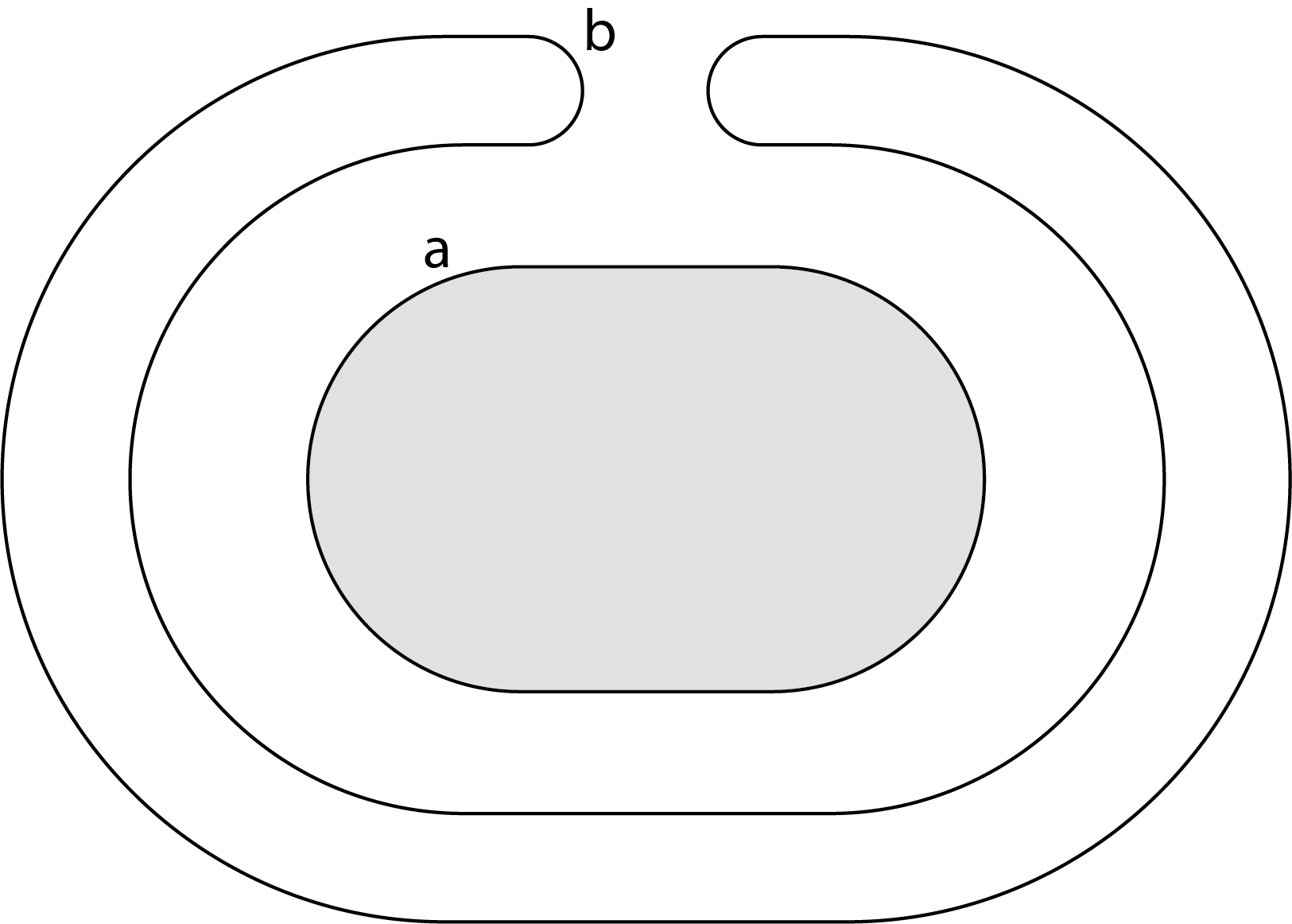}
\end{center}
\caption{Schematic diagram showing how a closed trajectory like that in
Fig.~\ref{f7}, which follows the border of the gray region, does not enclose
that region. The biggest trajectory inside the gray region is labeled {\bf a};
a trajectory just outside the border of the gray region is labeled {\bf b}.
Note that {\bf a} and {\bf b} are different. Outside the the gray region the
{\bf b} trajectory snaps and turns around. It snakes in one direction on a new
sheet of the Riemann surface and then snakes in the opposite direction on
another sheet, and thus does not enclose any part of the gray region.}
\label{f8}
\end{figure}

One surprising and nonintuitive result of the numerical study in Fig.~\ref{f7}
is that the trajectory that lies at the border but immediately {\it outside} the
gray Region $R_0$ in Fig.~\ref{f7} does {\it not} enclose the turning points
$t_0$ in $R_0$. It is important to understand how a trajectory that follows the
border of a region does not enclose that region. To assist in visualizing why
this is so, we provide the schematic diagram in Fig.~\ref{f8}. This figure shows
that a trajectory just outside the border of the gray region snaps and heads off
in one direction, turns around, heads off in another direction, turns around,
and then returns to its starting point without enclosing the gray region.

\begin{figure}
\begin{center}
\includegraphics[scale=0.32]{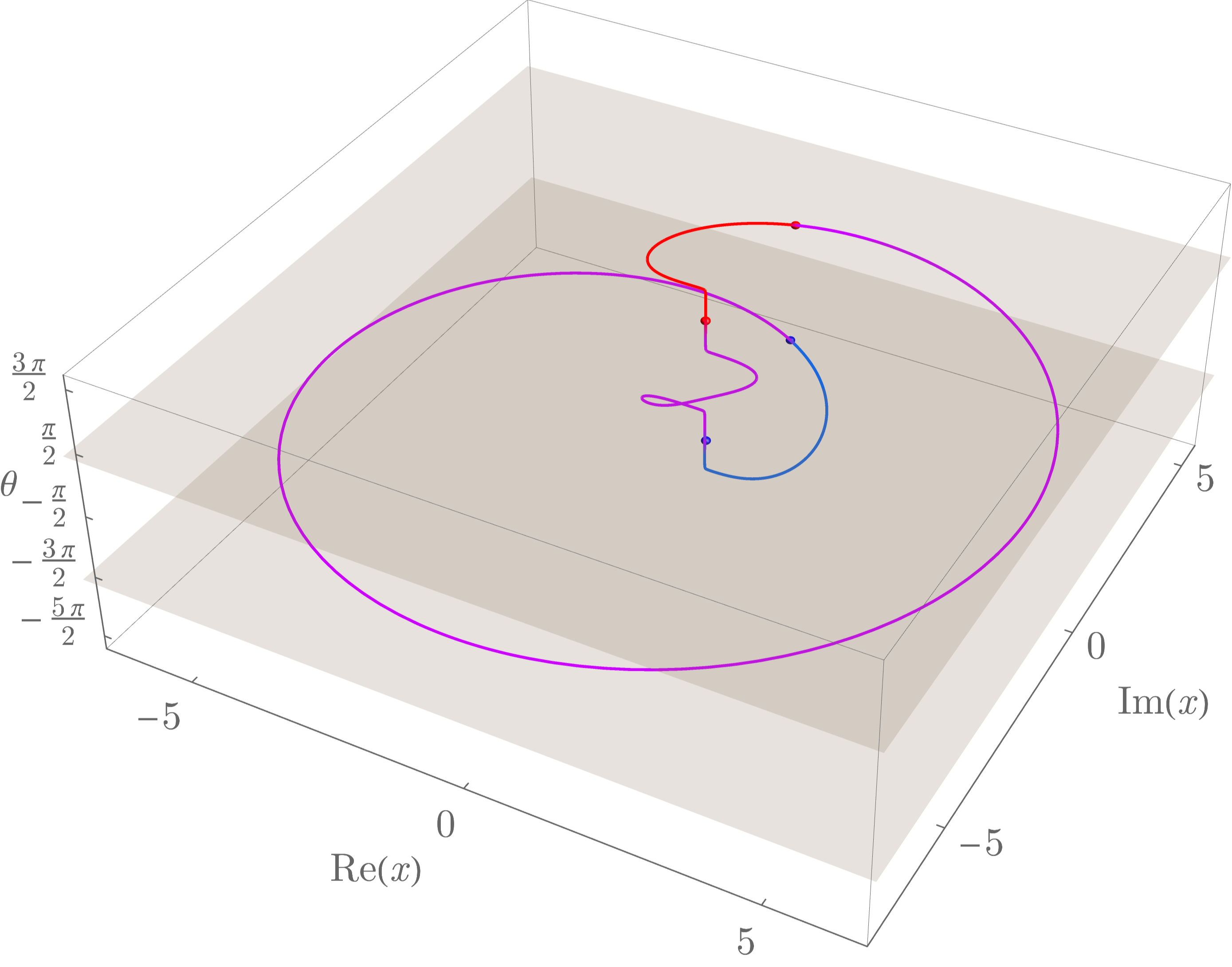}
\end{center}
\caption{[Color online] The trajectory in Fig.~\ref{f7} plotted in
three-dimensional perspective to make precise the generic features of the curve
{\bf b} in the schematic diagram in Fig.~\ref{f8}. At the red dots the
trajectory enters and leaves sheet 1 and at the blue dots the trajectory
enters and leaves sheet $-1$ on the Riemann surface. The trajectory begins
at $-0.68\,i$.}
\label{f9}
\end{figure}

Having seen the schematic diagram in Fig.~\ref{f8}, we display in Fig.~\ref{f9}
a three-dimensional view of the trajectory in Fig.~\ref{f7}. This figure
emphasizes that the trajectory crosses from sheet 0 (the principal sheet) up
onto sheet $+1$ and then down through sheet 0 and onto sheet $-1$ before
returning back to sheet 0 of the Riemann surface.

\begin{figure}
\begin{center}
\includegraphics[scale=0.27]{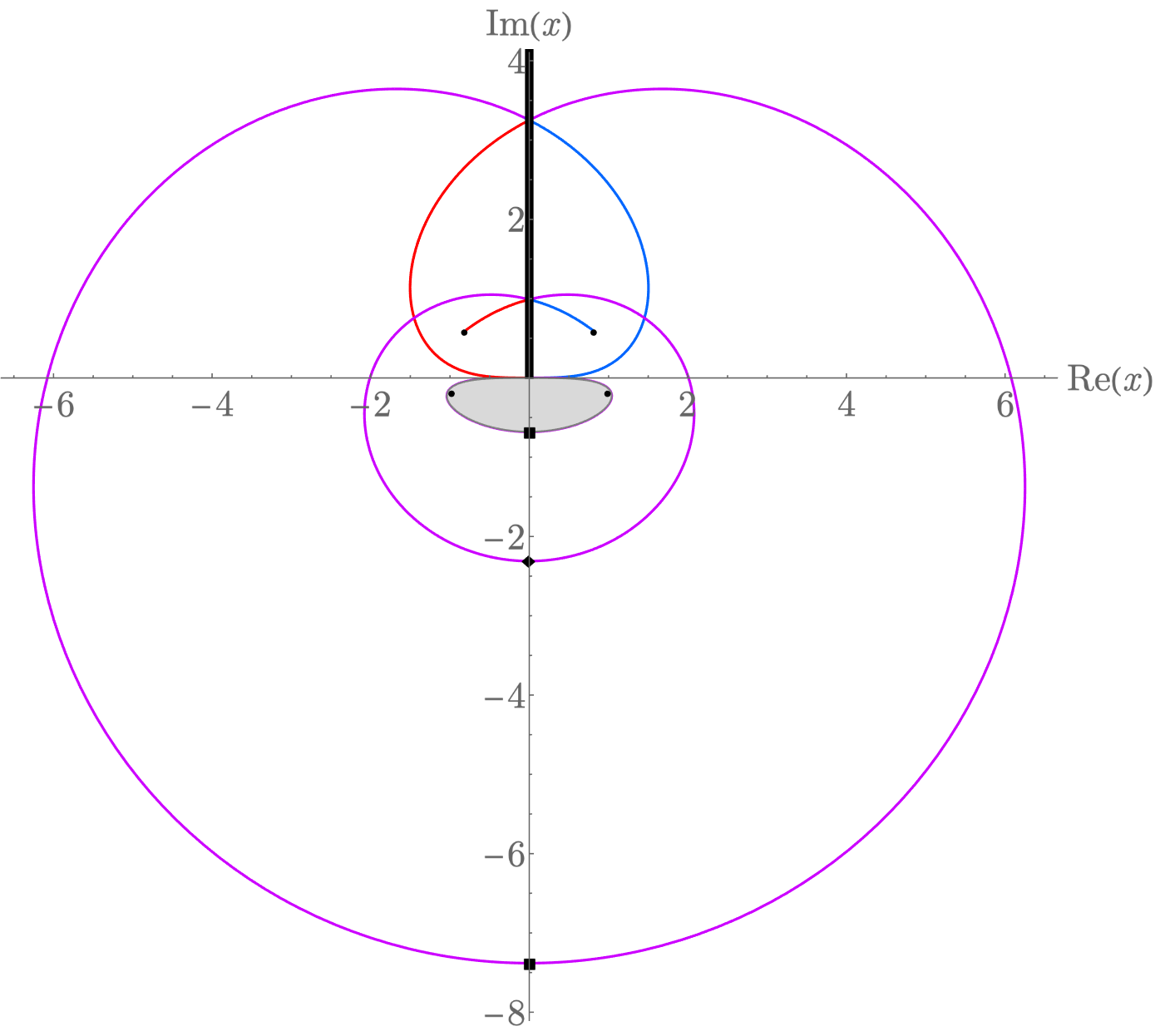}
\end{center}
\caption{[Color online] Same as Fig.~\ref{f7} but with one additional trajectory
that terminates at the two new turning points at $t_1=\pm 0.81697+0.57668,i$ in
the upper-half complex planes on sheets $-1$ and 1. This new trajectory begins
lower down the imaginary-$x$ axis at $s_1=-2.31061\,i$. As in Fig.~\ref{f7},
trajectories are colored purple on sheet 0, red on sheet 1, and blue on sheet
$-1$.} 
\label{f10}
\end{figure}

Like the special trajectory in Fig.~\ref{f4}, which connects the two turning
points on the principal sheet, there is also a special terminating trajectory
that connects the two new turning points $t_1$ indicated in Fig.~\ref{f7}, which
are located on sheets $+1$ and $-1$. This special trajectory is the complex
generalization of real oscillatory classical motion: Rather than going around
the turning points, the classical particle that follows this trajectory reaches
a turning point, stops, and then retraces its path as it moves away from the
turning point. This terminating trajectory, which is shown in Fig.~\ref{f10},
begins at $s_1=-2.31061\,i$.

All of the trajectories in Region $R_0$ are $\cPT$ symmetric and cross the
negative-imaginary axis on the principal sheet of the Riemann surface between
the origin and $x_0$. Similarly, all of the trajectories in Region $R_1$ are
$\cPT$ symmetric and cross the negative-imaginary axis on the principal sheet of
the Riemann surface between $x_0$ and $x_1$. This might lead us to conjecture
that there is an infinite tower of regions $R_n$ $(n=0,\,1,\,2,\,...$, where
$R_n$ contains the pair of turning points $t_n$ joined by the terminating
trajectory $s_n$, and one might guess that these regions cover the
negative-imaginary axis and that they are bounded by separatrices passing
through the critical points $x_n$. While this is a very neat and simple picture,
this hypothesis is completely wrong!

The actual situation is much more complicated. Only a {\it finite} number of
regions cover the entire negative-imaginary axis, and this number decreases to 1
as $\vep$ approaches 2 from below. The lower end of the lowest region extends to
$-i\infty$. Indeed, we see in Fig.~\ref{f5} that as $\vep\to2^-$, the critical
point $x_0$ reaches $-i\infty$.  For the case were are considering, namely,
$\vep=\frac1\pi$, the negative axis is completely covered by just four Regions:
$R_0$, $R_1$, $R_2$, and $R_5$, where the lower end of Region $R_5$ is at $-i
\infty$. Because the negative-imaginary axis is covered by these four regions,
it is not possible for trajectories associated with any other regions to cross
this axis, and thus all other trajectories are not $\cPT$ symmetric. 

It is possible that some of these broken $\cPT$-symmetric paths might approach
limit cycles in the complex plane, but we have not yet been found such a
trajectory. It is much more likely that non-$\cPT$-symmetric trajectories spiral
outward towards infinity as we have seen in Fig.~\ref{f1} (left panel).

We have found a particularly useful class of trajectories that help us to
understand the topology of complex classical orbits, and these are
asymptotically radial trajectories that terminate at infinity. To find such
trajectories we begin with (\ref{e4}), and study the asymptotic behavior of
$x(t)$ for large $|x|$. We introduce radial coordinates via
\begin{equation}
x(t)=r(t)e^{i\theta_0},
\label{e12}
\end{equation}
where $r(t)$ and $\theta_0$ are real and $r(t)>0$. In general, the polar angle
is a function of $t$, as we see in Fig.~\ref{f1} (left panel), but here we seek
{\it asymptotically radial} paths so we are interested in the simple case for
which $\theta_0$ is a constant. Substituting (\ref{e12}) into (\ref{e4}), we
obtain an ordinary differential equation for $r(t)$:
\begin{equation}
\dot{r}^2 e^{2i\theta_0}+4r^{2+\vep}e^{i\pi\vep/2}e^{(2+\vep)i\theta_0}=4.
\label{e13}
\end{equation}

If there is a solution for which $r\gg1$, we may neglect the 4 on the right side
of (\ref{e13}). We therefore have a two-term differential equation,
\begin{equation}
\dot{r}^2 \sim -4r^{2+\vep}e^{i\pi\vep/2}
e^{i\vep\theta_0}\quad(r\to\infty),
\label{e14}
\end{equation}
which we can solve. Since the left side of (\ref{e14}) is real and positive, the
right side must also be real and positive, so we obtain an algebraic equation
for the phase
\begin{equation}
1=-e^{i\pi\vep/2} e^{i\vep\theta_0},
\label{e15}
\end{equation}
and a separable asymptotic differential equation for the radial variable $r(t)$:
\begin{equation}
\dot{r}^2\sim 4r^{2+\vep}\quad(r\to\infty).
\label{e16}
\end{equation}

First, we solve (\ref{e16}). Remembering that both sides are positive, we take
the square root and get $\dot{r}\sim\pm 2r^{1+\vep/2}~~(r\to\infty)$. We then
{\it separate} the equation $\dot{r}r^{-1-\vep/2}\sim\pm2~~(r\to\infty)$ and
integrate both sides with respect to $t$: $r^{-\vep/2}\sim\pm\vep(C-t)~~(r\to
\infty)$, where we have divided by $-2$ and $C$ is a constant of integration.
Finally, we square this equation, invert it, and raise it to the power $1/\vep$:
\begin{equation}
r^{\vep}\sim[\vep^2(C-t)^2]^{-1/\vep}\quad(r\to\infty).
\label{e17}
\end{equation}
This shows that for all $\vep>0$ the particle reaches $r=\infty$ in {\bf finite
time}. Also we verify that the asymptotic approximation of neglecting the 4 on
the right side of (\ref{e13}) is consistent. Finally, we solve (\ref{e15}) to
find the angular locations of the asymptotic rays. The result is
\begin{equation}
\theta_0=-\frac{\pi}{2}+(2N-1)\frac{\pi}{\vep},
\label{e18}
\end{equation}
where $N=0,\,\pm1,\,\pm2,\,\ldots$ is an integer.

\begin{figure}
\begin{center}
\includegraphics[scale=0.23]{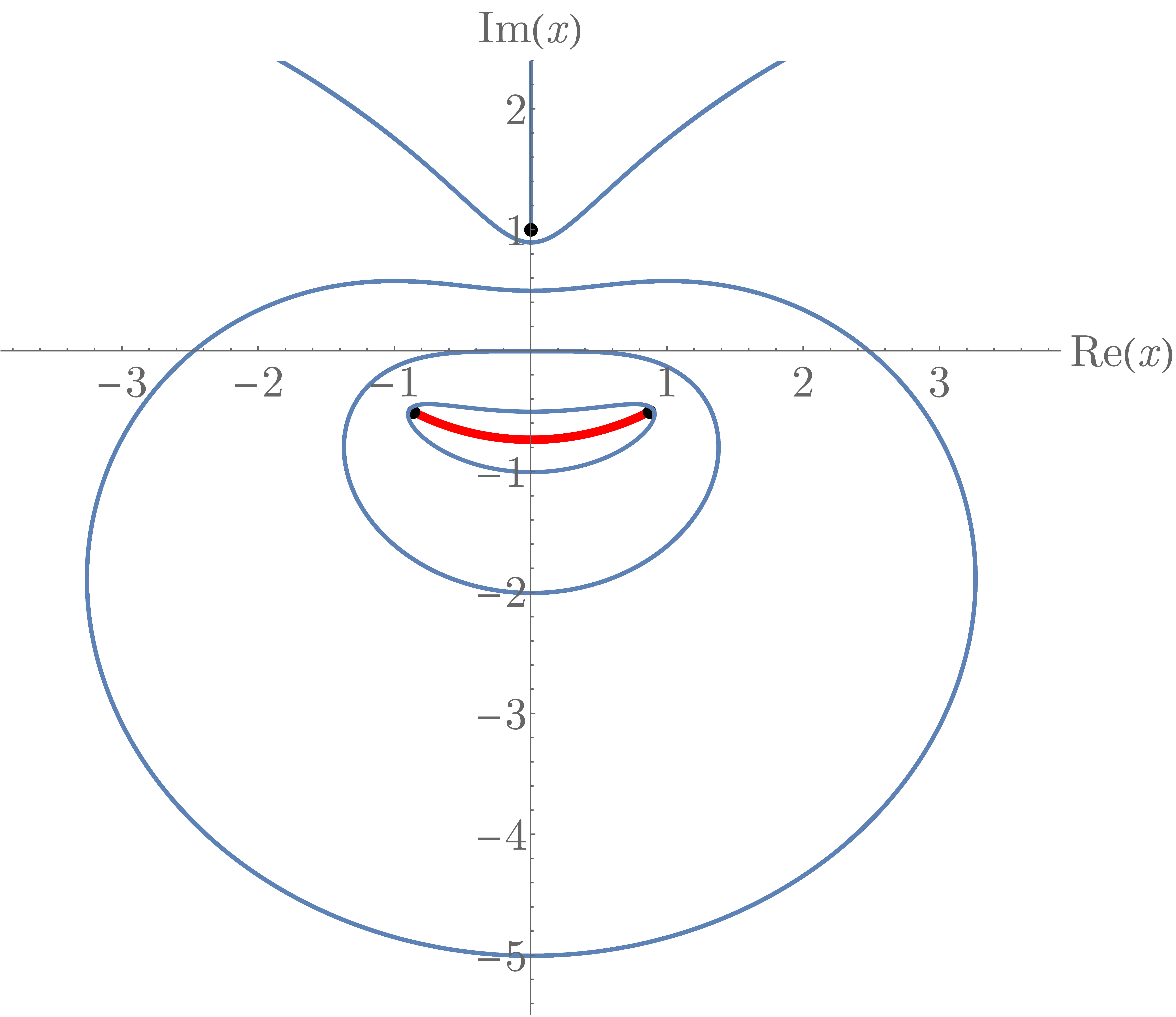}
\end{center}
\caption{[Color online] Complex trajectories for the simple case $\vep=1$. There
is only one sheet in the Riemann surface and three turning points (black dots).
Five trajectories are shown: (i) A two-ended terminating trajectory connecting
the $t_0$ pair of turning points in the lower-half plane; (ii) A single-ended
(semi-terminating) trajectory that begins at the turning point at $i$ and runs
off to $i\infty$ along the positive-imaginary axis; (iii) Three closed
trajectories, one beginning at $-i$, a second beginning at $-2i$ and passing
through the origin (as predicted in Fig.~\ref{f5}), and a third beginning at
$-5i$ and dipping to avoid the semi-terminating trajectory that begins at $i$.}
\label{f11}
\end{figure}

Here are a few simple cases: If $\vep=1$, then for $N=1$ the escape path (the
radial trajectory to infinity) is exactly at $\pi/2$. (See Fig.~\ref{f11}.) This
is a particularly interesting trajectory that terminates at {\it one} end.
The trajectory runs straight up the positive-imaginary axis from the turning
point at $i$ and bounces off the point at $i\infty$. If $\vep=2$, then the
escape path is located at $(N-1)\pi$. Thus, for $N=1$ and $N=0$ the escape paths
are at $0$ and $-\pi$ (that is, the positive-real and negative-real axes).

\begin{figure}
\begin{center}
\includegraphics[scale=0.55]{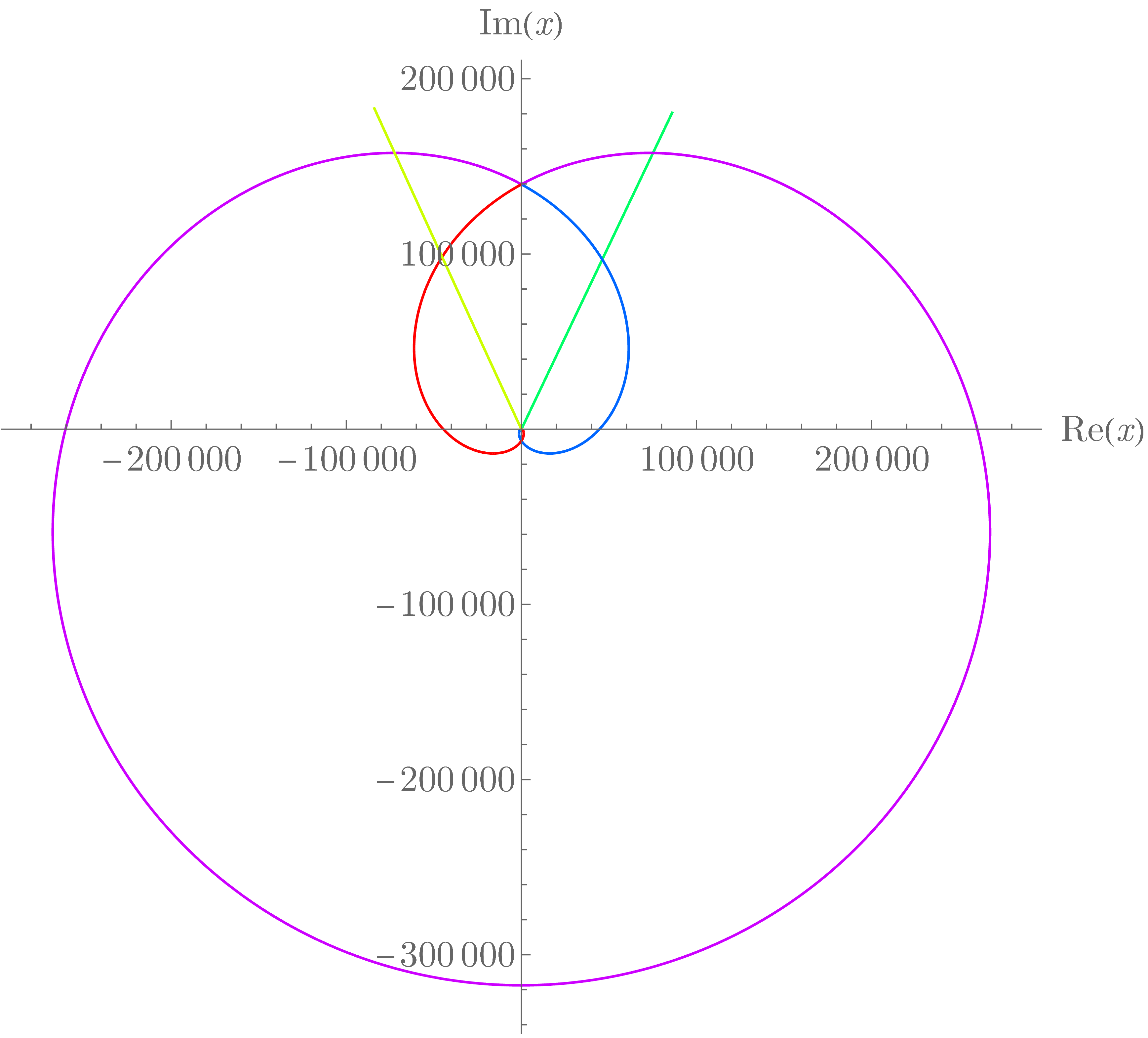}
\end{center}
\caption{[Color online]
A $\cPT$-symmetric five-sheeted separatrix path for $\vep=\frac{1}{\pi}$
corresponding the $N=0$ and $N=1$ cases in (\ref{e18}). The path becomes
asymptotically straight on sheets $\pm2$. The color scheme for the trajectory is
as follows: lime green on sheet $+2$, red on sheet $+1$, purple on sheet 0, blue
on sheet $-1$, and mint green on sheet $-2$.}
\label{f12}
\end{figure}

A more interesting and nontrivial asymptotically radial escape path for $\vep=
\frac1\pi$ is shown in Fig.~\ref{f12}. This path, which corresponds to $N=0$ and
$N=1$ in (\ref{e18}), lies on five sheets of the Riemann surface. The path is
$\cPT$ symmetric and approaches $\infty$ on sheets $\pm2$ of the Riemann
surface. This path is a separatrix trajectory between Regions $R_2$ and $R_5$.

\section{Behavior of complex trajectories for $\vep>2$}
\label{s4}
\begin{figure}
\begin{center}
\includegraphics[scale=0.55]{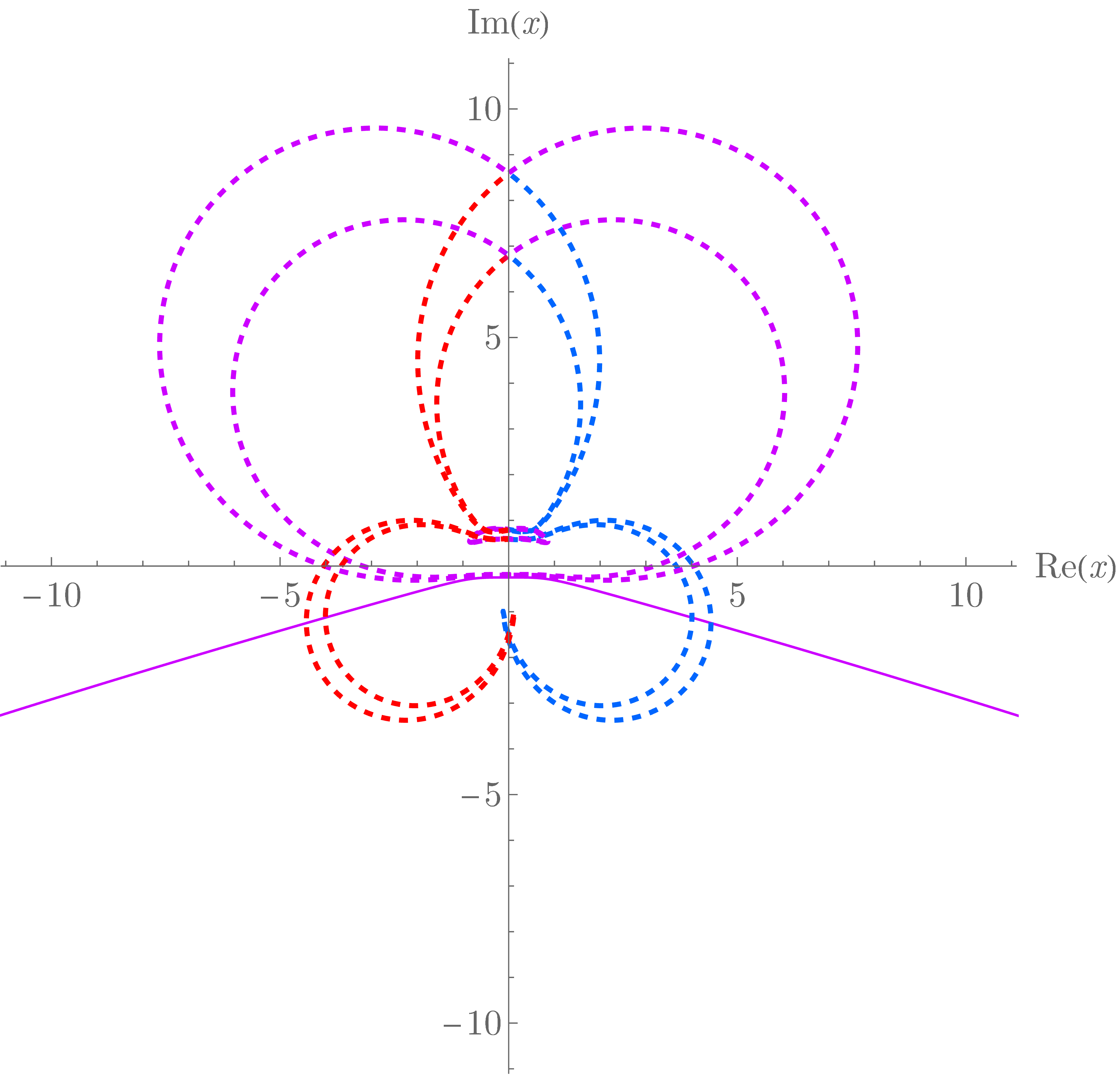}
\end{center}
\caption{[Color online]
Two complex trajectories for $\vep=1+\sqrt{2}$. One closed trajectory (solid
line) begins at $x_0=-0.25i$, which lies just below the upper edge of Region
$R_0$, and lies just inside the boundary Region $R_0$. This trajectory crosses
the negative-imaginary axis on the principal sheet again (not shown). A second
closed trajectory (dashed line) begins at $-0.2i$, which lies in the gap on the
negative-imaginary axis below the origin and above the upper edge of Region
$R_0$. This trajectory is $\cPT$ symmetric and lies inside Region $R_1$. Unlike
the situation for which $0<\vep<2$, Region $R_1$ lies {\it above} Region $R_0$.
The color scheme is the same as that used in Fig.~\ref{f12}.}
\label{f13}
\end{figure}

A strange remarkable change in the topology of the trajectories on Riemann
surface occurs as $\vep$ increases above 2. When $\vep>2$, the upper end of
Region $R_0$ on the negative-imaginary axis lies {\it below} the origin; a
narrow {\it gap} on the negative-imaginary axis opens up between the upper end
of Region $R_0$ and the origin. This gap is illustrated in Fig.~\ref{f13}, where
we have taken $\vep=1+\sqrt{2}$. A trajectory lying inside and very near the
upper edge of Region $R_0$ and below the gap begins at $-0.25i$ (solid line). A
second trajectory (dashed line) begins at $-0.2i$, which lies inside the gap.
This $\cPT$-symmetric trajectory lies inside Region $R_1$, which now lies {\it
above, and not below} Region $R_0$, unlike the configuration of regions for the
case $0<\vep<2$. Both trajectories are closed.

\begin{figure}
\begin{center}
\includegraphics[scale=0.30]{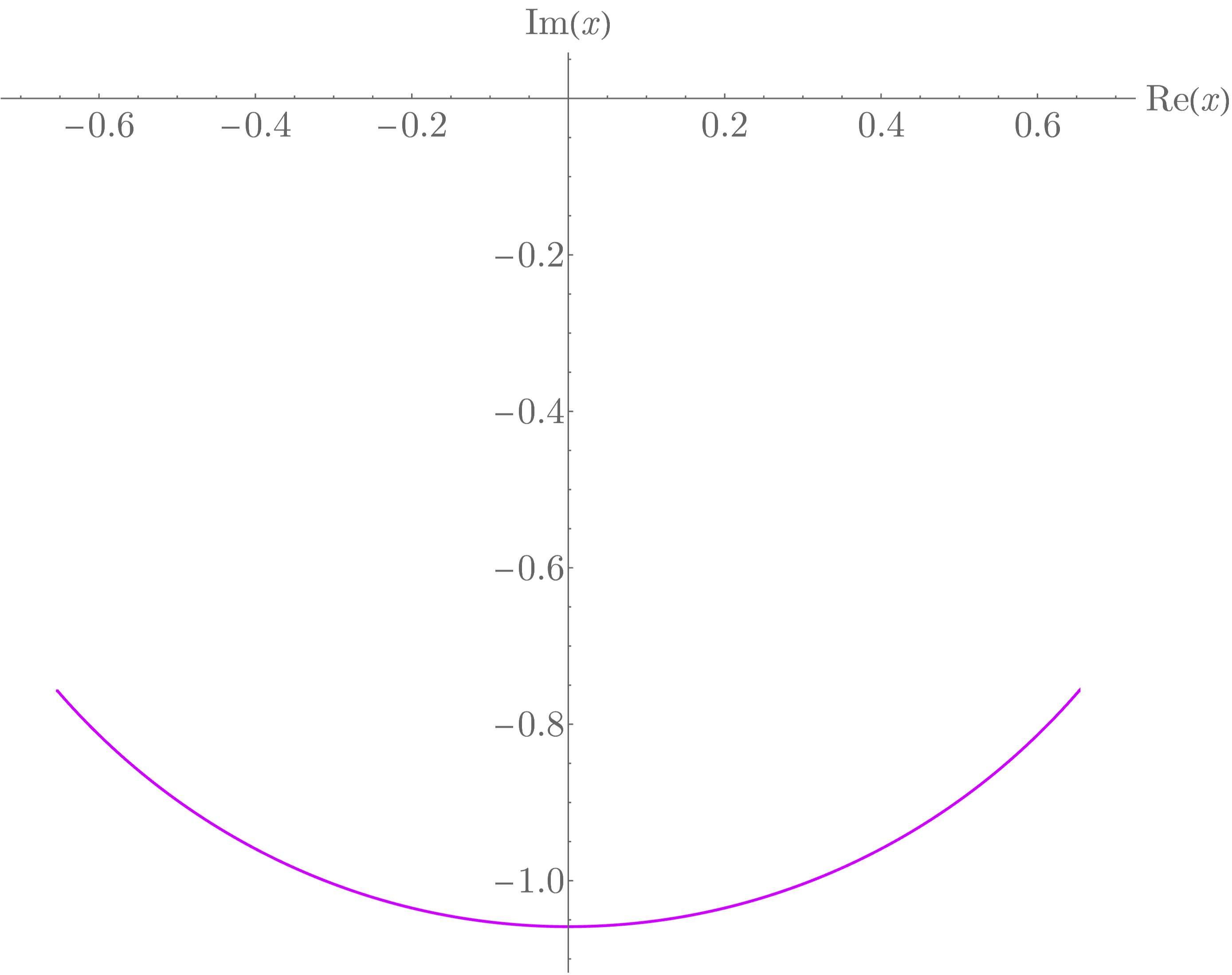}
\end{center}
\caption{[Color online]
Terminating path on the principal sheet for $\vep=\sqrt{2}+1$. This path
connects the $t_0$ pair of turning points and crosses the negative-imaginary 
axis at $-1.05872i$, which is consistent with Fig.~\ref{f6}. The color
scheme is the same as that used in Fig.~\ref{f12}.}
\label{f14}
\end{figure}

\begin{figure}
\begin{center}
\includegraphics[scale=0.29]{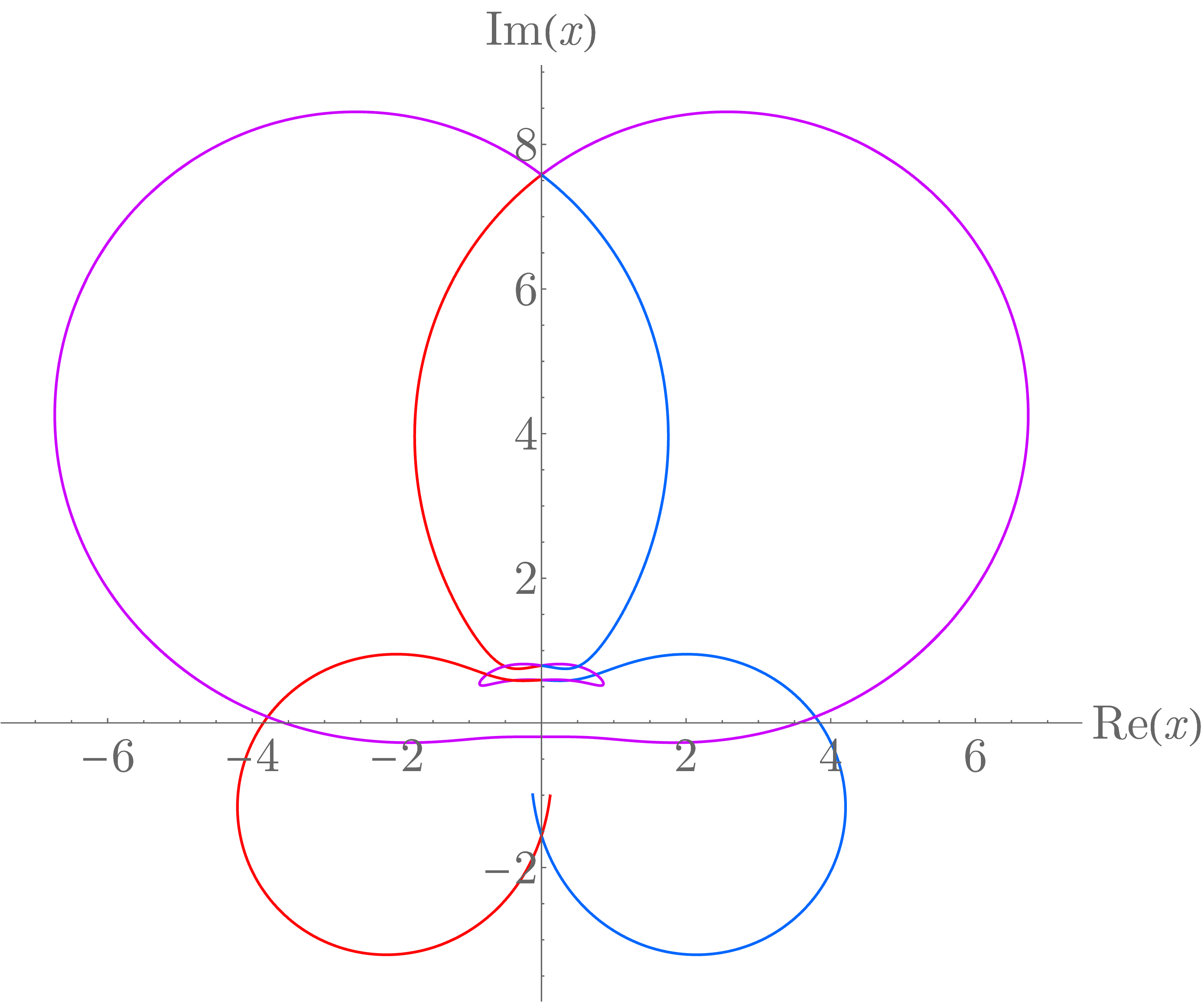}
\end{center}
\caption{[Color online]
Terminating path for $\vep=\sqrt{2}+1$ that connects the $t_4$ pair of turning
points and crosses the negative-imaginary axis on the principal sheet at
$-0.191947i$, which lies inside the gap. The color scheme is the same as
that used in Fig.~\ref{f12}.}
\label{f15}
\end{figure}

\begin{figure}
\begin{center}
\includegraphics[scale=0.30]{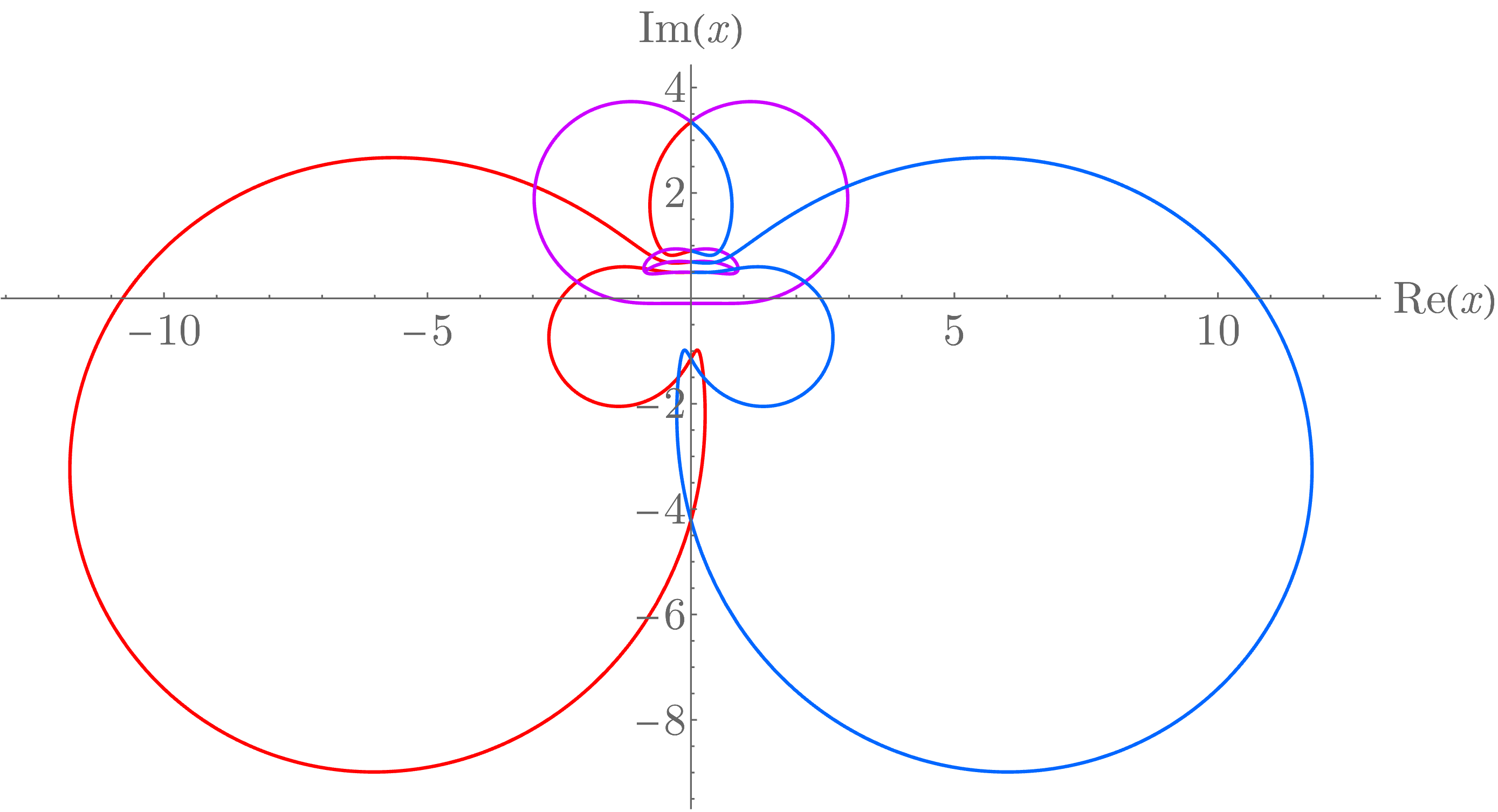}
\end{center}
\caption{[Color online]
Terminating path for $\vep=\sqrt{2}+1$ that connects the $t_1$ pair of turning 
points and crosses the negative-imaginary axis on the principal sheet at
$-0.0958837i$ inside the gap. The color scheme is the same as that used in
Fig.~\ref{f12}. This trajectory is slightly more complicated than that in
Fig.~\ref{f15}.}
\label{f16}
\end{figure}

\begin{figure}
\begin{center}
\includegraphics[scale=0.29]{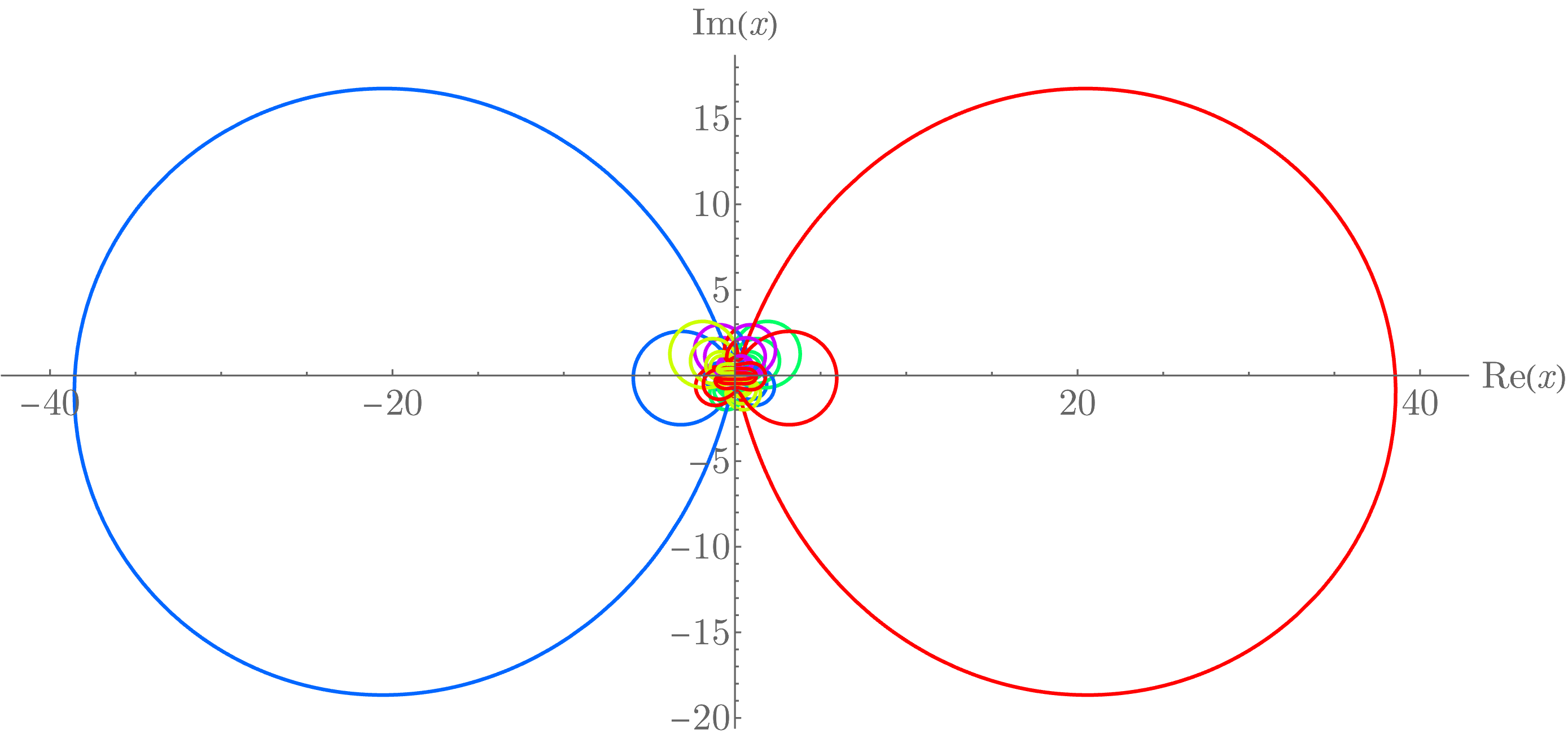}
\end{center}
\caption{[Color online]
Terminating path for $\vep=\sqrt{2}+1$ that connects the $t_7$ pair of turning
points and crosses the negative-imaginary axis on the principal sheet at
$-0.0469295i$ in the gap. The color scheme is the same as that used in
Fig.~\ref{f12}. This trajectory is much more complicated than that in
Fig.~\ref{f16} and visits 5 sheets in the Riemann surface.}
\label{f17}
\end{figure}

\begin{figure}
\begin{center}
\includegraphics[scale=0.33]{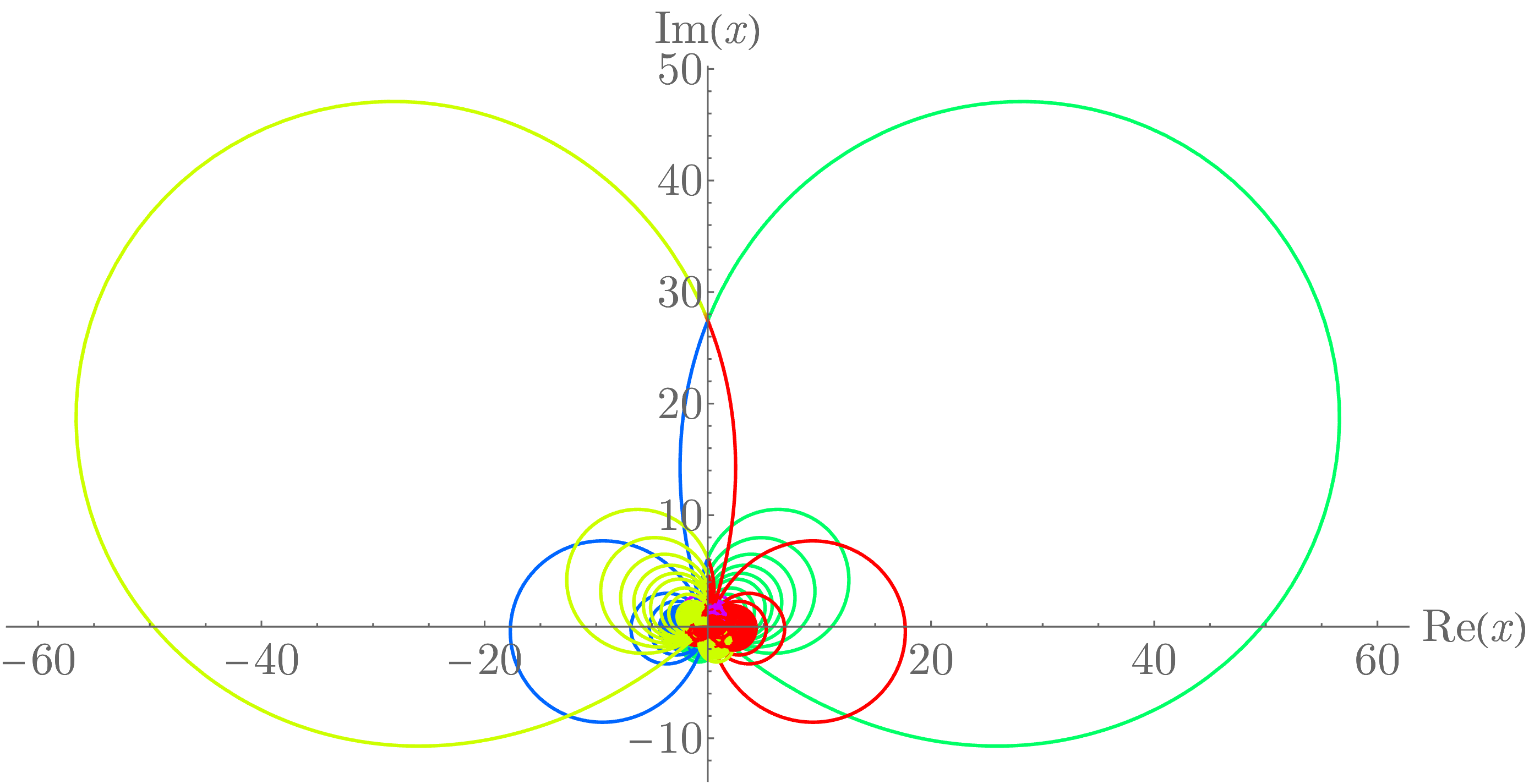}
\end{center}
\caption{[Color online]
Terminating path for $\vep=\sqrt{2}+1$ that connects the $t_5$ pair of turning
points and crosses the negative-imaginary axis on the principal sheet at
$-0.0312037i$ in the gap. The color scheme is the same as that used in
Fig.~\ref{f12}. This trajectory is much more complicated than that in
Fig.~\ref{f17}. The trajectory winds back and forth repeatedly as it searches
for the very tiny subgap through with it must pass in order to cross through the
negative-imaginary axis on the principal sheet.}
\label{f18}
\end{figure}

\begin{figure}
\begin{center}
\includegraphics[scale=0.33]{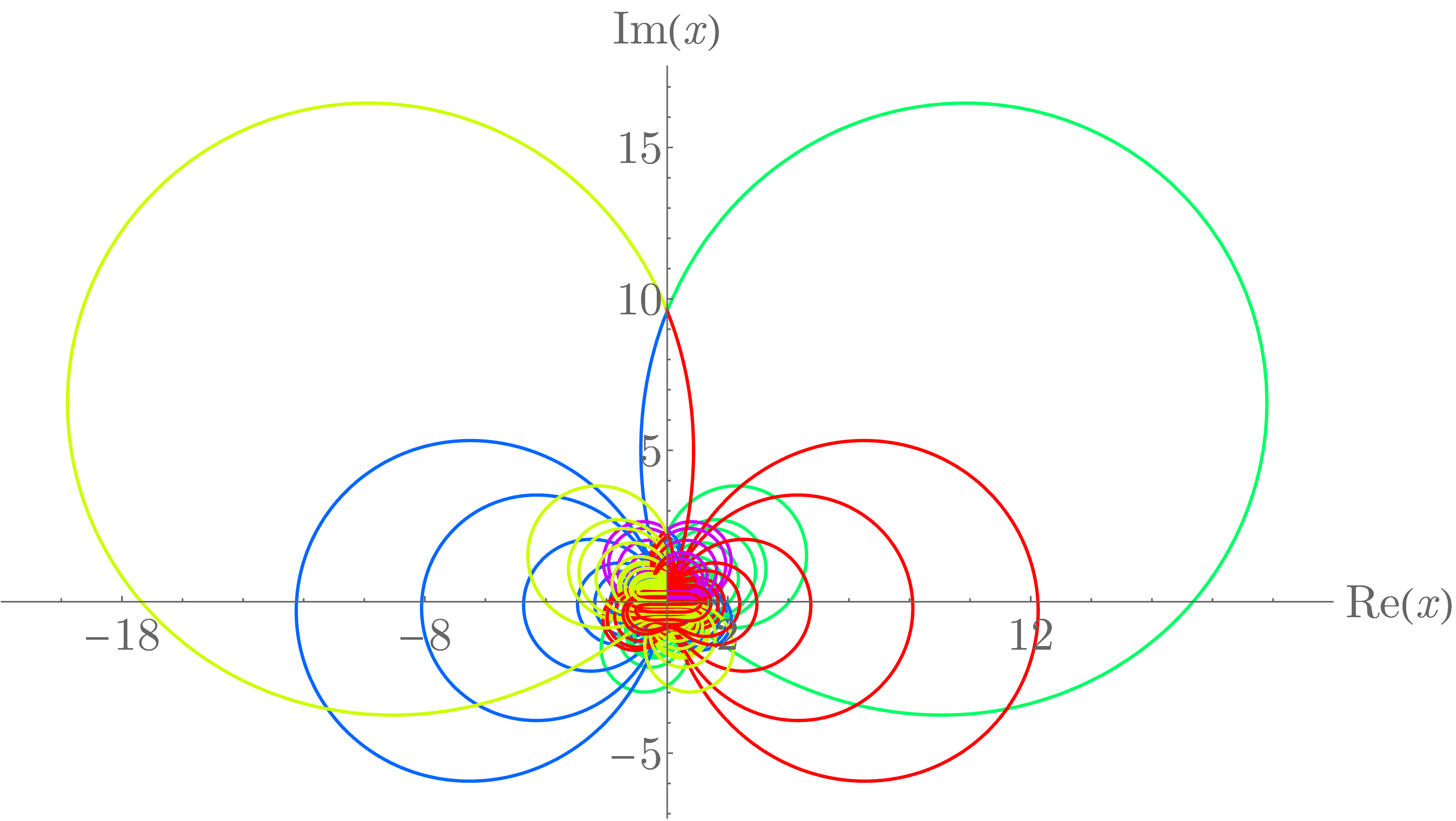}
\end{center}
\caption{[Color online]
Terminating path for $\vep=\sqrt{2}+1$ that connects the $t_3$ pair of turning
points and crosses the negative-imaginary axis on the principal sheet at
$-0.0167618i$ in the gap. The color scheme is the same as that used in
Fig.~\ref{f12}. This trajectory is even more complicated than that in
Fig.~\ref{f18}.}
\label{f19}
\end{figure}

\begin{figure}
\begin{center}
\includegraphics[scale=0.33]{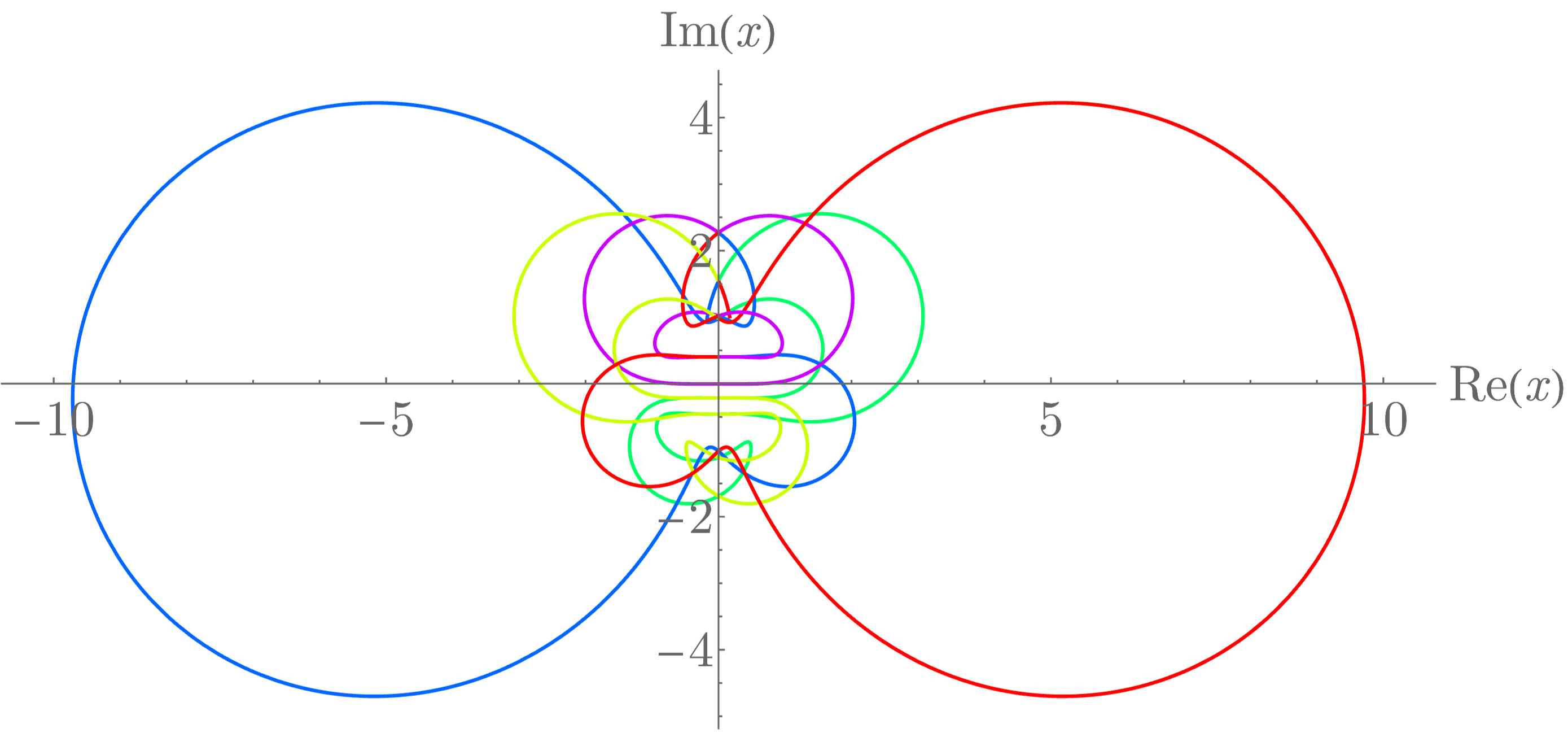}
\end{center}
\caption{[Color online]
Terminating path for $\vep=\sqrt{2}+1$ that connects the $t_6$ pair of turning
points and crosses the negative-imaginary axis on the principal sheet at
$-0.00378715i$ in the gap. The color scheme is the same as that used in
Fig.~\ref{f12}. It is interesting that this terminating trajectory has a
simpler structure than those in the previous figures. We do not have an
explanation for this.}
\label{f20}
\end{figure}

\begin{figure}
\begin{center}
\includegraphics[scale=0.25]{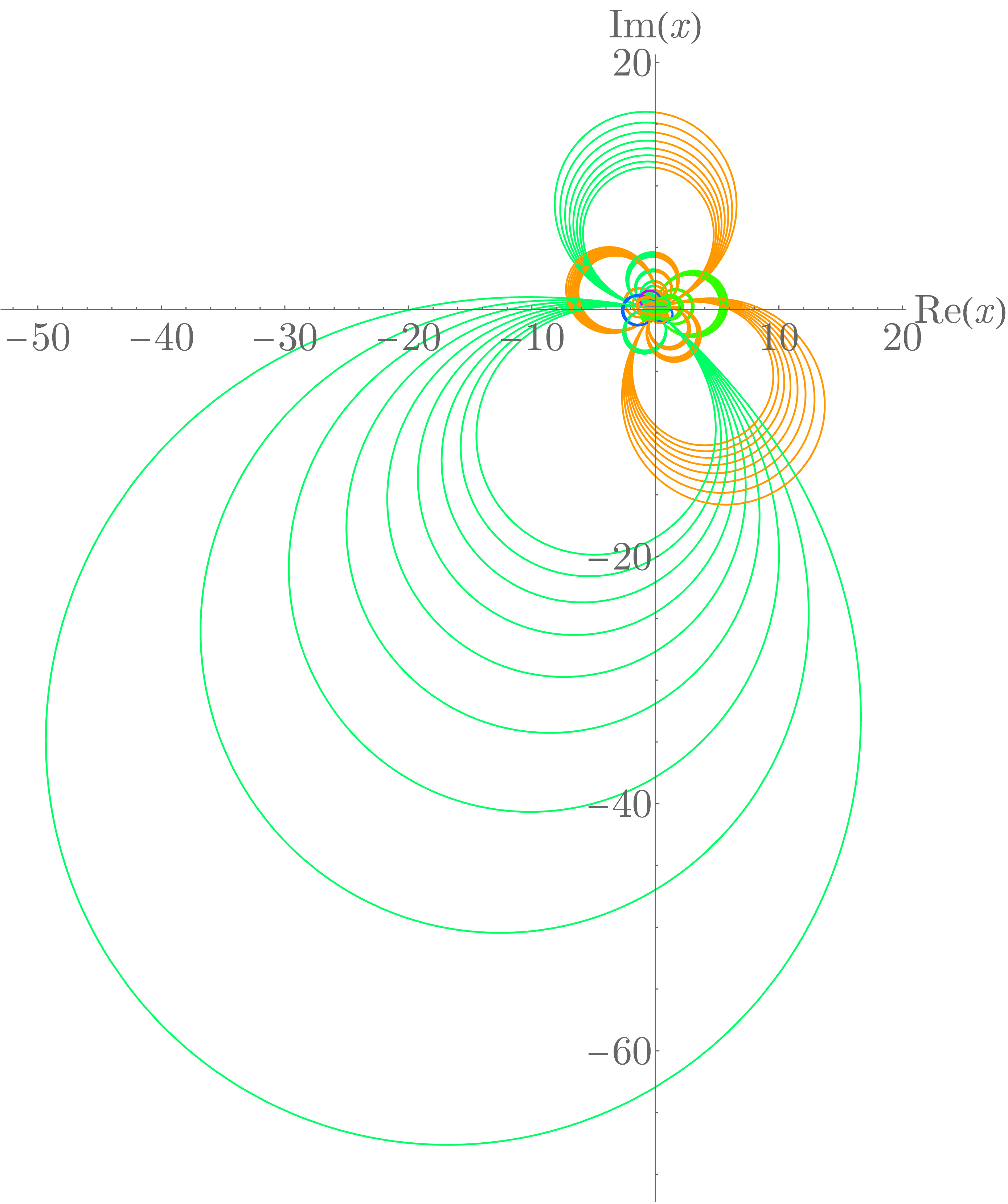}
\end{center}
\caption{[Color online]
Complex trajectory for $\vep=\sqrt{2}+1$ that begins at one of the $t_2$ turning
points, namely, the turning point on sheet $-1$. The path is unable to find a
way to cross through the gap on the negative-imaginary axis on the principal
sheet and thus it is not $\cPT$ symmetric. The path has a remarkable structure
in which it dips briefly onto sheet 0 and then oscillates between sheets $-2$
and $-3$ as it gradually spirals outward to complex infinity. The color scheme
is the same as that used in Fig.~\ref{f12} except that the trajectory is colored
brown on sheet $-3$.}
\label{f21}
\end{figure}

\begin{figure}
\begin{center}
\includegraphics[scale=.22]{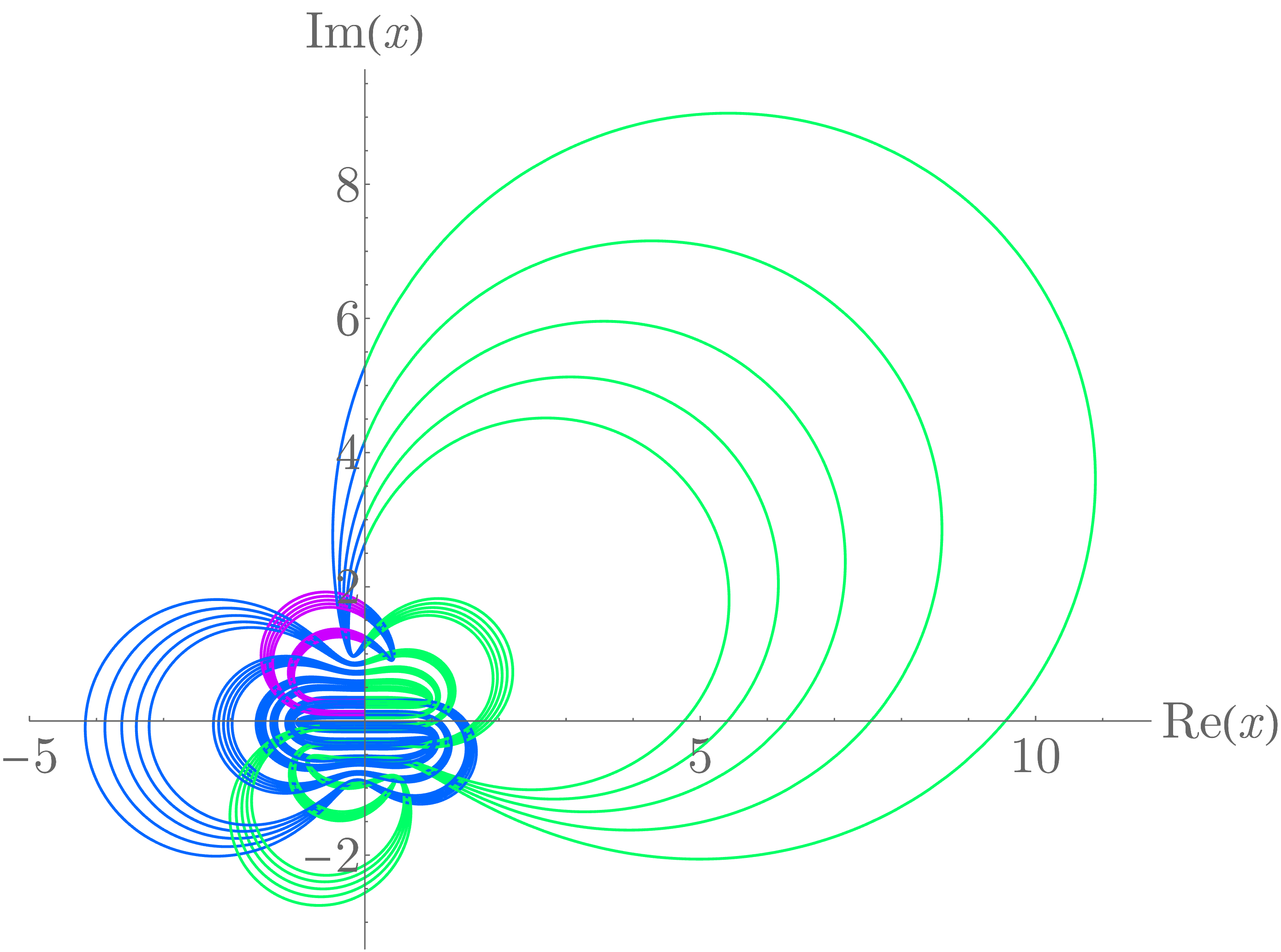}
\hspace{0.05cm}
\includegraphics[scale=.22]{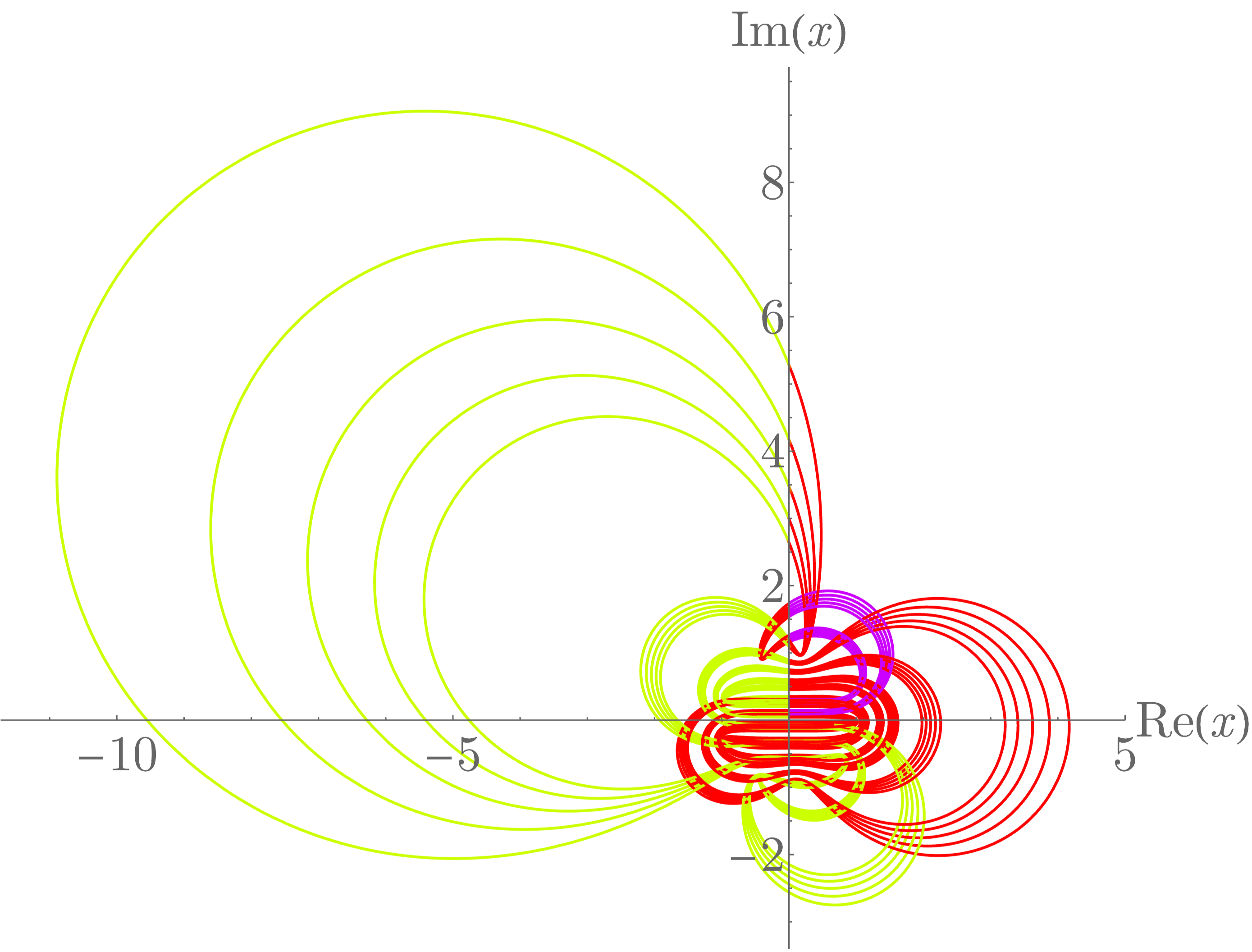}
\end{center}
\caption{[Color online]
Complex trajectories for $\vep=\sqrt{2}+1$ that begin at each of the $t_8$
turning points. Neither path is unable to cross through the gap on the
negative-imaginary axis on the principal sheet and thus they are not $\cPT$
symmetric. However, they are $\cPT$ reflections of one another. The color scheme
is the same as that used in Fig.~\ref{f12}.}
\label{f22}
\end{figure}

The gap on the negative-imaginary axis has a truly remarkable substructure. The
gap consists of many, and perhaps infinitely many, subintervals through which 
trajectories that originate on sheets $+1,\,+2,\,+3,\,\ldots$ can pass in order 
to enter sheets $-1,\,-2,\,-3,\,\ldots$ and thereby become $\cPT$ symmetric.
This means that the higher $\cPT$-symmetric regions develop extremely narrow
throats. To illustrate this substructure we have investigated the terminating
trajectories for $\vep=1+\sqrt{2}$ belonging to Regions $R_0,\ldots,\,R_8$ to
determine the points at which these paths pass through the gap. These crossing
points are listed in Table \ref{t1} and the terminating trajectories are
plotted in Figs.~\ref{f14} -- \ref{f22}. The trajectories that originate at the
$n=2$ and $n=8$ turning points do not cross the negative imaginary axis on the
principal sheet and instead spiral out to infinity. These trajectories are shown
in Figs. \ref{f21} and \ref{f22}. Note that the trajectories become increasingly
complicated as the crossing point gets closer to the top of the gap (at the
origin).

\begin{table}[h!]
\begin{center}
\begin{tabular}{|c|c|}
\hline
$n$ & Crossing points $s_n$ on negative-imaginary axis\\
\hline
0 & $-1.05872i$ \\
4 & $-0.191947i$ \\
1 & $-0.0958837i$ \\
7 & $-0.0469295i$ \\
5 & $-0.0312037i$ \\
3 & $-0.0167618i$ \\
6 & $-0.00378715i$ \\
\hline
\end{tabular}
\end{center}
\caption{\label{t1} 
Points $s_n$ on the negative-imaginary axis on the principal sheet at which
terminating trajectories (that is, trajectories that connect the $t_n$ pair of
turning points $t_n$) for $\sqrt{2}+1$ go from positive sheets to negative
sheets on the Riemann surface. The terminating trajectory in Region $R_0$
passes through the point $s_0$, which is near $-i$, and this is consistent with
the graph in Fig.~\ref{f6}. The remaining points $s_n$ lie inside the narrow gap
that lies between the origin and $-0.25i$, but these points come in an
unexpected order: 4, 1, 7, 5, 3, 6. There are {\it no} terminating trajectories
for $n=2$ and $n=8$. For these cases the paths that begin at one of the turning
points $t_2$ and $t_8$ are unable to pass through the gap; these paths are not
$\cPT$ symmetric and instead spiral outward to infinity.}
\end{table}

\section{Conclusions and conjectures regarding future studies}
\label{s5}
We have opened a Pandora's box; this study reveals the immense complexity of the
classical paths for the quantum-mechanical Hamiltonian in (\ref{e1}), much of
which had not been discovered in previous studies. Perhaps, the most surprising
discovery is the narrow gap on the negative-imaginary axis on the principal
sheet that opens up as $\vep$ increases above 2. Evidently, the point $\vep=2$
is an elaborate and unexpected singularity in the classical-mechanical theory. 

The presence of the narrow gap on the negative-imaginary axis of the principal
sheet of the Riemann surface of complex trajectories suggests two possibilities.
For those regions $R_n$ containing $\cPT$-symmetric classical trajectories, all
of these trajectories must pass through an extremely narrow subgap. Therefore,
these regions must be barbell-shaped with an extremely narrow throat. Mapping
the boundaries of these regions is not an easy numerical project but it is one
of our objectives and it is currently under way. All other regions contain
trajectories that are not $\cPT$ symmetric and these regions must come in
$\cPT$-symmetric pairs. One of our long-term objectives is to construct a clear
map of the Riemann surface. We expect to discover further structures. For
example, we may find that some of the region boundaries have fractal structure
\cite{R26}.

Most importantly, we believe that there may be more singular values of $\vep$ at
which additional gaps on the negative-imaginary axis appear. We conjecture that
a new gap appears at $\vep=6$, $\vep=10$, $\vep=14$, and so on, and that these
gaps have quantum-mechanical analogs that were discovered in Ref.~\cite{R25}. In
the quantum theory these are the points at which new families of entirely real
spectra appear as a consequence of the emergence of new $\cPT$-symmetric pairs
of Stokes sectors.

\vspace{.5cm}

CMB is supported by the Alexander von Humboldt Foundation, the Simons
Foundation, and the UK Engineering and Physical Sciences Research Council.
Mathematica was used to perform the numerical calculations in this paper.


\vspace{1.0cm}
\begin{thebibliography}{999} 
\bibitem{R1} C.~M.~Bender and S.~Boettcher, {\it Phys.~Rev.~Lett.}~{\bf 80},
5243 (1998).

\bibitem{R2} C.~M.~Bender, S.~Boettcher, and P.~N.~Meisinger, {\it
J.~Math.~Phys.}~{\bf 40}, 2201 (1999).

\bibitem{R3} C.~M.~Bender, {\it Reps.~Prog.~Phys.}~{\bf 70}, 947 (2007).

\bibitem{R4} P.~E.~Dorey, C.~Dunning, and R.~Tateo, {\it J.~Phys.~A:
Math.~Gen.}~{\bf 34}, L391 (2001) and {\bf 34}, 5679 (2001).

\bibitem{R5} P.~E.~Dorey, C.~Dunning, and R.~Tateo, {\it J.~Phys.~A:
Math.~Gen.}~{\bf 40}, R205 (2007).

\bibitem{R6} C. M. Bender {\it et al}, {\it PT symmetry: In quantum and
classical systems} (World Scientific, Singapore, 2019).

\bibitem{R7} H. F. Jones and J. Mateo, {\it Phys. Rev. D} {\bf 73}, 085002
(2006).

\bibitem{R8} C. M. Bender, D. C. Brody, J.-H. Chen, H. F. Jones, K. A. Milton,
and M. C. Ogilvie, {\it Phys. Rev. D} {\bf 74}, 025016 (2006).

\bibitem{R9} C. M. Bender and D. W. Hook, {\it J. Phys. A: Math. Theor.} {\bf
41}, 244005 (2008).

\bibitem{R10} C. M. Bender, D. C. Brody, and H. F. Jones, {\it Phys. Rev. Lett.}
{\bf 93}, 251601 (2004).

\bibitem{R11} C.~M.~Bender, D.~C.~Brody, and D.~W.~Hook, {\it J. Phys. A:
Math. Theor.}~{\bf 41}, 352003 (2008).

\bibitem{R12} J.~Rubinstein, P.~Sternberg, and Q.~Ma, {\it Phys.~Rev.~Lett.}
{\bf 99}, 167003 (2007).

\bibitem{R13} A.~Guo, G.~J.~Salamo, D.~Duchesne, R.~Morandotti,
M.~Volatier-Ravat, V.~Aimez, G. A. Siviloglou, and D.~N.~Christodoulides,
{\it Phys.~Rev.~Lett.}~{\bf 103}, 093902 (2009).

\bibitem{R14} C.~E.~R\"uter, K.~G.~Makris, R.~El-Ganainy, D.~N.~Christodoulides,
M.~Segev, and D.~Kip, {\it Nat.~Phys.}~{\bf 6}, 192 (2010).

\bibitem{R15} K.~F.~Zhao, M.~Schaden, and Z.~Wu, {\it Phys.~Rev.}~A {\bf 81},
042903 (2040).

\bibitem{R16} Y.~D.~Chong, L.~Ge, and A.~D.~Stone, {\it Phys. Rev. Lett.}~{\bf
106}, 093902 (2011).

\bibitem{R17} Z.~Lin, H.~Ramezani, T.~Eichelkraut, T.~Kottos, H.~Cao, and
D.~N.~Christodoulides, {\it Phys. Rev.~Lett.}~{\bf 106}, 213901 (2011).

\bibitem{R18} J.~Schindler, A.~Li, M.~C.~Zheng, F.~M.~Ellis, T.~Kottos, {\it
Phys.~Rev.}~A {\bf 84}, 040101 (2011).

\bibitem{R19} S.~Bittner, B.~Dietz, U.~Guenther, H.~L.~Harney, M.~Miski-Oglu,
A.~Richter, and F.~Schaefer, {\it Phys.~Rev.~Lett.} {\bf 108}, 024101 (2012).

\bibitem{R20} N. M. Chtchelkatchev, A. A. Golubov, T. I. Baturina, and V. M.
Vinokur, {\it Phys. Rev. Lett.} {\bf 109}, 150405 (2012).

\bibitem{R21} B. Peng, S. K. Ozdemir, F. Lei, F. Monifi, M. Gianfreda, G. L.
Long, S. Fan, F. Nori, C. M. Bender, L. Yang, {\it Nat. Phys.} {\bf 10}, 394
(2014).

\bibitem{R22} S. Assawaworrarit, X. Yu, and S. Fan, {\it Nat.} {\it 546}, 387
(2017). 

\bibitem{Z1} C. M. Bender, D. D. Holm, and D. W. Hook, {\it J. Phys. A: Math.
Theor.} {\bf 40}, f81 (2006).

\bibitem{Z2} C. M. Bender, J.-H. Chen, D. W. Darg, and K. A. Milton, {\it J.
Phys. A: Math. Theor.} {\bf 39}, 4219 (2006).

\bibitem{Z3} C. M. Bender and D. W. Darg, {\it J. Math. Phys.} {\bf 48},
042703 (2007).

\bibitem{Z4} T. Arpornthip and C. M. Bender, {\it Pramana} {\bf 73}, 259
(2009).

\bibitem{Z5} C. M. Bender, D. W. Hook, and K. S. Kooner, {\it J.~Phys.~A: Math.
Theor.} {\bf 43}, 165201 (2010).

\bibitem{Z6} A. G. Anderson and C. M. Bender, {\it J. Phys. A: Math. Theor.}
{\bf 45}, 455101 (2012).

\bibitem{Z7} C. M. Bender and D. W. Hook, {\it J. Phys. A: Math. Theor.} {\bf
44}, 372001 (2011).

\bibitem{Z8} C. M. Bender and D. W. Hook, {\it Phys. Rev. A} {\bf 86}, 022113
(2012).

\bibitem{Z9} C. M. Bender, D. W. Hook, N. E. Mavromatos, and S. Sarker, {\it
J. Phys. A: Math. and Theor.} {\bf 49}, 45lt01 (2016).

\bibitem{R23} C. M. Bender and D. W. Hook, {\it Stud. Appl. Math.} {\bf 133},
318 (2014). 

\bibitem{R24} C. M. Bender, S. Boettcher, H. F. Jones, and V. M. Savage,
{\it J. Phys. A: Math. Gen.} {\bf 32}, 6771 (1999).

\bibitem{R25} C. M. Bender and S. P. Klevansky, {\it Phys. Rev. Lett.} {\bf
105}, 031601 (2010).

\bibitem{R26} C. M. Bender and J. P. Vinson, J. Math. Phys. {\bf 37}, 4103
(1996).

\end{thebibliography}
\end{document}